\documentclass[12pt]{article}

\usepackage{graphicx}
\usepackage{amsmath,marvosym,enumerate,multirow,amsfonts}

\usepackage{verbatim}
\usepackage{fancyvrb}
\usepackage{algorithm}
\usepackage{algpseudocode}
\usepackage{mathrsfs}
\usepackage{xcolor, tikz, fancybox, morefloats}
\usepackage{lineno}

\usepackage{lineno}

\newtheorem{example}{Example}

\definecolor{lightgray}{gray}{0.9}

\newenvironment{evb}{\VerbatimEnvironment%
    \noindent\begin{Sbox}\footnotesize
        \begin{minipage}{\dimexpr\linewidth-10\fboxsep-5\fboxrule}
          \begin{Verbatim}
}{%
      \end{Verbatim}%
      \end{minipage}%
      \end{Sbox}%
      \fcolorbox{black}{lightgray}{\TheSbox}%
}

\providecommand{\keywords}[1]{\textbf{\textit{Keywords:}} #1}

\begin{document}

\title{Development of A Scalable Platform for Large-scale Reservoir Simulations on Parallel Computers}

%% use optional labels to link authors explicitly to addresses:
%% \author[label1,label2]{<author name>}
%% \address[label1]{<address>}
%% \address[label2]{<address>}

\author{Hui Liu\thanks{Authors to whom correspondence may be addressed. Email addresses:
hui.sc.liu@gmail.com}, Kun Wang, Bo Yang, and Zhangxin Chen \\
Department of Chemical and Petroleum Engineering \\
University of Calgary \\
Calgary, Alberta, Canada, T2N 1N4}

\date{}
\maketitle

\begin{abstract}
This paper presents our work on designing a platform for large-scale reservoir simulations. Detailed components, such as grid and linear solver, and data structures are introduced, which can serve as a guide to parallel reservoir simulations and other parallel applications.  The main objective of platform is to support implementation of various parallel reservoir simulators on distributed-memory parallel systems, where MPI (Message Passing Interface) is employed for communications among computation nodes.  It provides structured grid due to its simplicity and cell-centered data is applied for each cell.  The platform has a distributed matrix and vector module and a map module. The matrix and vector module is the base of our parallel linear systems. The map connects grid and linear system modules, which defines various mappings between grid and linear systems.  Commonly-used Krylov subspace linear solvers are implemented, including the restarted GMRES method and the BiCGSTAB method. It also has an interface to a parallel algebraic multigrid solver, BoomerAMG from HYPRE. Parallel general-purpose preconditioners and special preconditioners for reservoir simulations are also developed. Various data structures are designed, such as grid, cell, data, linear solver and preconditioner, and some key default parameters are presented in this paper.  The numerical experiments show that our platform has excellent scalability and it can simulate giant reservoir models with hundreds of millions of grid cells using thousands of CPU cores.

\bigskip
\noindent \keywords{platform, reservoir simulation, parallel computing, algorithm, data structure}
\end{abstract}

%%
%% Start line numbering here if you want
%%
%\setpagewiselinenumbers
%\linenumbers

\section{Introduction}
Nowadays, various operation processes have been developed to enhance oil recovery by the oil and gas industry.
Their numerical simulations are becoming more and more complicated.
In the meantime, geological models from reservoirs are more and more complex, and they are also heterogenous.
Models with millions of grid cells are usually employed to obtain high resolution results.
Numerical simulations may take days or even longer to complete one run using regular workstations.
The long simulation time could be a problem to reservoir engineers,
since dozens of simulations may be required to find optimal operations.
Fast computational methods and reservoir simulators should be investigated.

Reservoir simulations have been studied for decades and various models and methods have been developed
by researchers, including black oil model, compositional model, thermal model and related topics.
Kaarstad et al. \cite{PS-Kaa} studied oil-water model and they implemented
a reservoir simulator that could solve problems with up to one million grid cells.
Rutledge et al. \cite{PS-Rut} developed a
compositional simulator for massive SIMD computers, which employed the IMPES (implicit pressure-explicit saturation) method.
Killough et al. \cite{PS-Kil} implemented a compositional simulator for distributed-memory parallel systems.
Killough et al. also used the locally refined grids in their parallel simulator to improve accuracy \cite{PS-Kil2}.
Dogru and his group \cite{PS-Dogru1, PS-Dogru2} developed a parallel simulator,
which was capable of simulating reservoir models with one billion grid cells.
Zhang et al. developed a platform for adaptive finite element and adaptive finite
volume methods, which has been applied to CFD, Maxwell equation, material, electronic structures,
biology and reservoir simulations \cite{phg, phg-quad, kwang}, and a black oil simulator using discontinuous
Galerkin method has been reported \cite{kwang}.
For many reservoir simulations, especially black oil simulation,
most of the simulation time is spent on the solution of linear systems
and it is well-known that the key of accelerating
linear solvers is to develop efficient preconditioners.
Many preconditioner methods have been applied to reservoir simulations, including point-wise and block-wise
incomplete factorization (ILU) methods for general linear systems \cite{ILU1}, domain decomposition methods
\cite{RAS}, constrained pressure residual (CPR) methods for the black oil
model, compositional model and extended black oil models \cite{CPR-old,CPR-cao},
multi-stage methods \cite{Study-Two-Stage}, multiple level preconditioners
\cite{mlp} and fast auxiliary space preconditioners (FASP) \cite{FASP}.

This paper presents our work on developing a parallel platform
for large-scale reservoir simulations on parallel systems and designing various data structures.
The platform is implemented using C and MPI (Message Passing Interface).
MPI is a standardized message-passing system designed to work on
a wide varieties of parallel system and it is employed to handle communications among computation nodes.
The platform provides structured grid, cell-centered data, linear solvers, preconditioners,
distributed matrices and vectors, visualization, parallel input and output through MPI-IO,
key words parsing and well modeling modules.
Finite difference methods and finite volume methods are supported.
The load balancing module is crucial for parallel computing \cite{HSFC, PaMETIS, pzhang1}.
The load balancing module is completed by ParMETIS \cite{PaMETIS} and the Hilbert
space-filling curve (HSFC) method \cite{HSFC}.
The ParMETIS is a graph partitioning package
using topological information of a grid, and the HSFC partitioning method
is an in-house partitioning method, which serves as the default partitioner.
Commonly used Krylov subspace solvers and algebraic multigrid (AMG) solvers are
implemented, including the restarted GMRES solver, BiCGSTAB solver \cite{Krylov}, and classic AMG solvers \cite{HYPRE2}.
General preconditioners, including ILU(k), ILUT, domain decomposition \cite{RAS} and AMG \cite{HYPRE2},
and special preconditioners, including CPR-like preconditioners, are implemented.
Detailed designs and key parameters are presented.
Numerical experiments show that our platform is capable of calculating problems with hundreds of millions of grid cells
and it has excellent scalability on distributed-memory parallel computers.

\section{Data Types}

Data structures in this paper use C language style.
Figure \ref{fig-ds-dt} presents our basic data types for integer, float-point, boolean and character types
and MPI-related data types used by MPI. These data types
are defined automatically by \texttt{configure},
which is generated by \verb|autoconf|, \verb|m4| and \texttt{BASH} scripts.

\begin{figure}[!htb]
\centering
\begin{evb}
/* float */
#if USE_LONG_DOUBLE
typedef long double              FLOAT;
#else
typedef double                   FLOAT;
#endif

/* integer */
#if USE_LONG_LONG
typedef signed long long int     INT;
#elif USE_LONG
typedef signed long int          INT;
#else
typedef signed int               INT;
#endif

typedef char                     CHAR;
typedef FLOAT                    COORD[3];
#undef  TRUE
#define TRUE                     (1)
#undef  FALSE
#define FALSE                    (0)
typedef int                      BOOLEAN;

/* MPI type */
#if USE_LONG_DOUBLE
#define PRSI_MPI_FLOAT           MPI_LONG_DOUBLE
#else
#define PRSI_MPI_FLOAT           MPI_DOUBLE
#endif

#if USE_LONG_LONG
#define PRSI_MPI_INT             MPI_LONG_LONG_INT
#elif USE_LONG
#define PRSI_MPI_INT             MPI_LONG
#else
#define PRSI_MPI_INT             MPI_INT
#endif

#define PRSI_MPI_CHAR            MPI_CHAR
#define PRSI_MPI_BOOLEAN         MPI_INT

/* CSR format */
typedef struct mat_csr_t_
{
    INT   num_rows;
    INT   num_cols;
    INT   num_nonzeros;
    INT   *Ap;
    INT   *Aj;
    FLOAT *Ax;

} mat_csr_t;
\end{evb}
\caption{Basic data types}
\label{fig-ds-dt}
\end{figure}

The default type for floating-point number is \verb|double|. If long double is enabled, its type
is \verb|long double|. Macro \verb|USE_LONG_DOUBLE| is defined to define derived types, such
as \verb|PRSI_MPI_FLOAT| and \verb|COORD|.
The default integer type is \verb|long int|. Two macros, \verb|USE_LONG_LONG| and \verb|USE_LONG|,
are defined to control integer types and derived data types, such as MPI types.
\verb|BOOLEAN| is for boolean type and \verb|CHAR| is for string.
\verb|mat_csr_t| is basic data type for CSR matrix.

\section{Grid}

A traditional reservoir can be described as $\Omega = [x_1, x_2] \times [y_1, y_2] \times [z_1, z_2]$.
If the domain $\Omega$ are divided into $n_x$, $n_y$ and $n_z$ intervals in the $x$, $y$ and $z$ directions,
then the grid has $N_g = n_x \times n_y \times n_z$ cells.
Each cell is a hexahedron. An interior cell has
six neighbors and each boundary cell may have three, four or five neighbors, depending on its location.
A structured grid can be uniform or non-uniform. The structured grids support finite difference methods, finite volume
methods and finite element methods, and they have been
widely used by commercial reservoir simulators.

Each cell has a unique global index.
Its default index is calculated as
\begin{equation}
C_{(i, j, k)} = n_x * n_y * k + n_x * j + i,
\end{equation}
which is numbered from the bottom layer of a reservoir to the top layer of the reservoir.
Here $i$, $j$ and $k$ are the integer coordinates of the cell
in the $x$, $y$ and $z$ directions, respectively.
Another numbering style used by most reservoir simulators is
\begin{equation}
C_{(i, j, k)} = n_x * n_y * (n_z - k) + n_x * j + i,
\end{equation}
which is numbered from top layer to bottom layer.

\begin{figure}[!htb]
\centering
\begin{evb}
typedef struct CELL_
{
#if !USE_LESS_MEMORY
    COORD      ctrd;                    /* centroid coordinate */
    FLOAT      area[6];                 /* area of each face */
    FLOAT      vol;                     /* volume */
#endif

    void       *nb[6];                  /* neighbours */
    INT        vert[8];                 /* local index of vertices */
    INT        index;                   /* local index */
    INT        idx[3];                  /* index in x, y and z direction */
    INT        regn;                    /* region mark */

    USHORT     bdry_type[6];            /* boundary type */
    USHORT     type;                    /* cell type */

} CELL;
\end{evb}
\caption{Data structure of CELL}
\label{fig-ds-cell}
\end{figure}

The data structure of a cell, which is defined as \verb|CELL|, is shown by Figure \ref{fig-ds-cell}.
This data structure defines cell-related information, such as
centroid coordinate (\verb|ctrd|), area of all faces (\verb|area|) and volume (\verb|vol|),
six neighbors (\verb|nb|), local index of each vertex (\verb|vert|), local index on an MPI process (\verb|index|),
global index in three directions (\verb|idx|), region mark (\verb|regn|),
boundary type of each face (\verb|bdry_type|) and cell type (\verb|type|).
Some members are optional. For example,
if we would like to use less memory, we can remove \verb|ctrd|, \verb|area| and \verb|vol|
by setting \verb|USE_LESS_MEMORY| to
some positive integer, such as \verb|1|.
In reservoir simulations, each cell represents a portion of reservoir and they have
similar properties, such as porosity, pressure, temperature, water saturation and oil saturation.

\begin{figure}[!htb]
\centering
\begin{evb}[fontsize=\small]
typedef struct RNEIGH_
{
    INT         gidx;
    INT         lidx;
    int         rank;

} RNEIGH;
\end{evb}
\caption{Data structure of remote neighbor}
\label{fig-ds-rnh}
\end{figure}

When a neighbor of a cell is in another MPI process, communication is required when accessing neighbor information. The
data structure \verb|RNEIGH| stores remote cell information, such as its global cell index (\verb|gidx|), its local
index (\verb|lidx|), and its MPI rank (\verb|rank|). The member \verb|rank| defines message sender and receiver during
communication.

\begin{figure}[!htb]
\centering
\begin{evb}[fontsize=\small]
typedef struct GRID_
{
    COORD           *vert;               /* coordinates of each vertex */
    CELL            *cell;
    INT             *num_cells;          /* number of cells in each process */
    RNEIGH          *rnghr;              /* remote neighbours */
    USHORT          *type_vert;          /* vert types */
    INT             *L2Gmap_vert;        /* Local to global map of vertices */
    INT             *L2Gmap_cell;        /* Local to global map of cell indices */

    FLOAT           lif;                 /* Load imbalance factor */
    INT             nregns;              /* number of region marks */
    INT             nverts;              /* number of vertices in the subgrid */
    INT             nfaces;              /* nuber of faces */
    INT             ncells;              /* number of cell indices in the subgrid */
    INT             nrngbr;              /* number of remote neighbours */

    INT             nfaces_remote;       /* equals to number of remote neighbours */
    INT             nverts_global;       /* number of vertices in the global grid */
    INT             nfaces_global;       /* number of vertices in the global grid */
    INT             ncells_global;       /* number of cells in the global grid */

    FLOAT           bbox[3][2];          /* bounding box */
    INT             ncx, ncy, ncz;       /* grid size in x, y, z directions */
    FLOAT           *vx, *vy, *vz;       /* partition of x, y, and z directions */
    BOOLEAN         uniform;             /* uniform in each direction or not */

    /* Well, (nprocs - 1)-th process owns all wells  */
    WELL            **well;
    WELL_CINFO      *well_cinfo;
    CELL            **perf_cell;        /* pointer to cell which has perferation */
    INT             nperfs;             /* number of perferations */
    INT             nperfs_global;
    INT             nwells_global;      /* number of wells */
    BOOLEAN         well_assembled;
    BOOLEAN         destroy_well;       /* if grid or simulator destroy wells */

    MPI_Comm        comm;
    int             rank;
    int             nprocs;

} GRID;
\end{evb}
\caption{Data structure of GRID}
\label{fig-ds-grid}
\end{figure}

Data structure for structured grids, \verb|GRID|, is presented in Figure \ref{fig-ds-grid},
which stores the coordinates of each vertex (\verb|vert|), vertex indices (\verb|L2Gmap_vert|),
cell indices (\verb|L2Gmap_cell|), distribution of cells (\verb|num_cells|) in each MPI process,
a mapping between the global index of a cell and its local index, well data and MPI info. In each MPI process,
a portion of a grid is stored, and only vertices and faces belong to these cells
are stored. Space cost of a grid in each MPI process is proportional to grid size (number of cells).

\subsection{Grid Partitioning}

Let $\mathbb{G}$ be the structured grid, which is distributed in $N_p$ MPI tasks,
\begin{equation}
\mathbb{G} = \{C_1, C_2, \cdots, C_{N_g}\},
\end{equation}
where $C_i$ is the $i$-th cell of $\mathbb{G}$. Each task owns a subset of $\mathbb{G}$, $\mathbb{G}_i$, which
satisfies the following conditions:
\begin{equation}
 \left\{
 \begin{aligned}
& \mathbb{G}_i \neq \emptyset \ (i = 1, \cdots, N_p) \\
& \mathbb{G}_i \cap \mathbb{G}_j = \emptyset \ (i \neq j) \\
& \cup \mathbb{G}_i = \mathbb{G} \ (i = 1, \cdots, N_p).
\end{aligned}
 \right.
\end{equation}
A cell belongs to some sub-grid, and its neighboring cells may belong to different sub-grids.

Each cell has similar calculations. Therefore, it is seasonable to assume that each cell has the
same amount of calculations. The workload of each MPI task can be modeled by the size
of its sub-grid, $|\mathbb{G}_i|$, or simply the number of grid cells in the sub-grid.
When discretizing reservoir models, information from neighboring
cells is always required.
For any sub-grid, its communication volume is determined by its remote neighboring cells,
which can be modeled by dual graph: we see a cell as a vertex of a graph, and if two cells are
neighbors, there exists an edge between these two cells.

The goal of grid partitioning is that each MPI task has equal workload (number of cells)
and the communications are minimized. Graph methods are ideal tools,
such as spectral methods \cite{nc2,nc9}, multilevel methods \cite{nc10,nc11,nc12}
and diffusive methods. Several graph partitioning packages have been implemented
and available publicly, such as ParMETIS, which is widely applied in parallel computing.
It is also used for matrix reordering.
The geometry information based methods are also efficient, including recursive coordinate
bisection method, recursive inertial bisection method and the space-filling curve methods \cite{HSFC}.

\subsection{Space-filling Partitioning Method}
This section introduces space-filling curve, Hilbert order, algorithm for generating
Hilbert order and space-filling curve partitioning method.

\subsubsection{Space-filling Curves}

Space-filling curves are those curves that fill an entire $n$-dimensional unit hypercube,
which were proposed by Peano in 1890 and popularized by Hilbert later.

\begin{figure}[!htb]
    \centering
    \includegraphics[width=.8\linewidth]{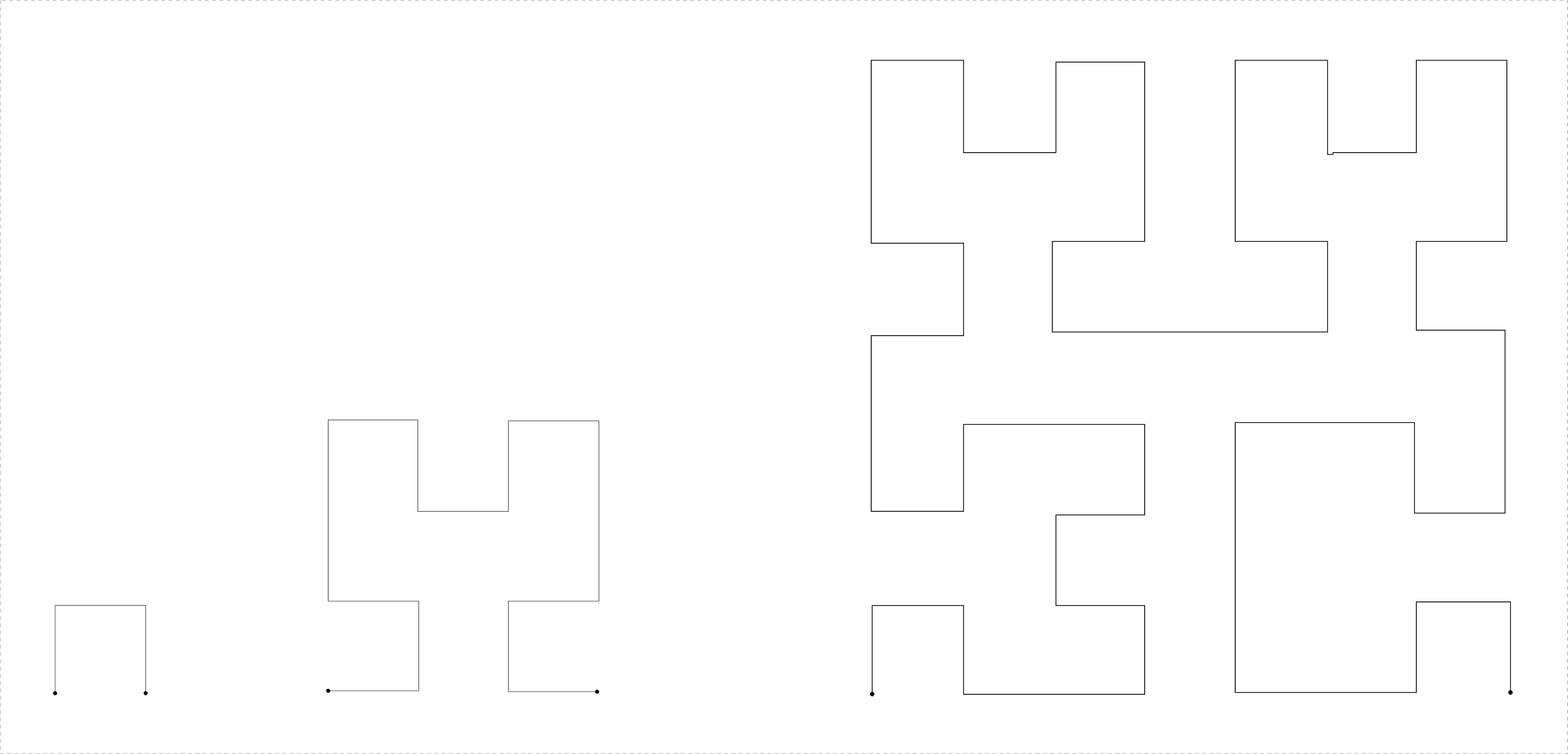}
    \caption{Hilbert space-filling curves, two dimensions, levels 1, 2 and 3}
    \label{fig-hilbert}
\end{figure}

\begin{figure}[!htb]
\begin{center}
\includegraphics[width=0.5 \hsize,angle=270]{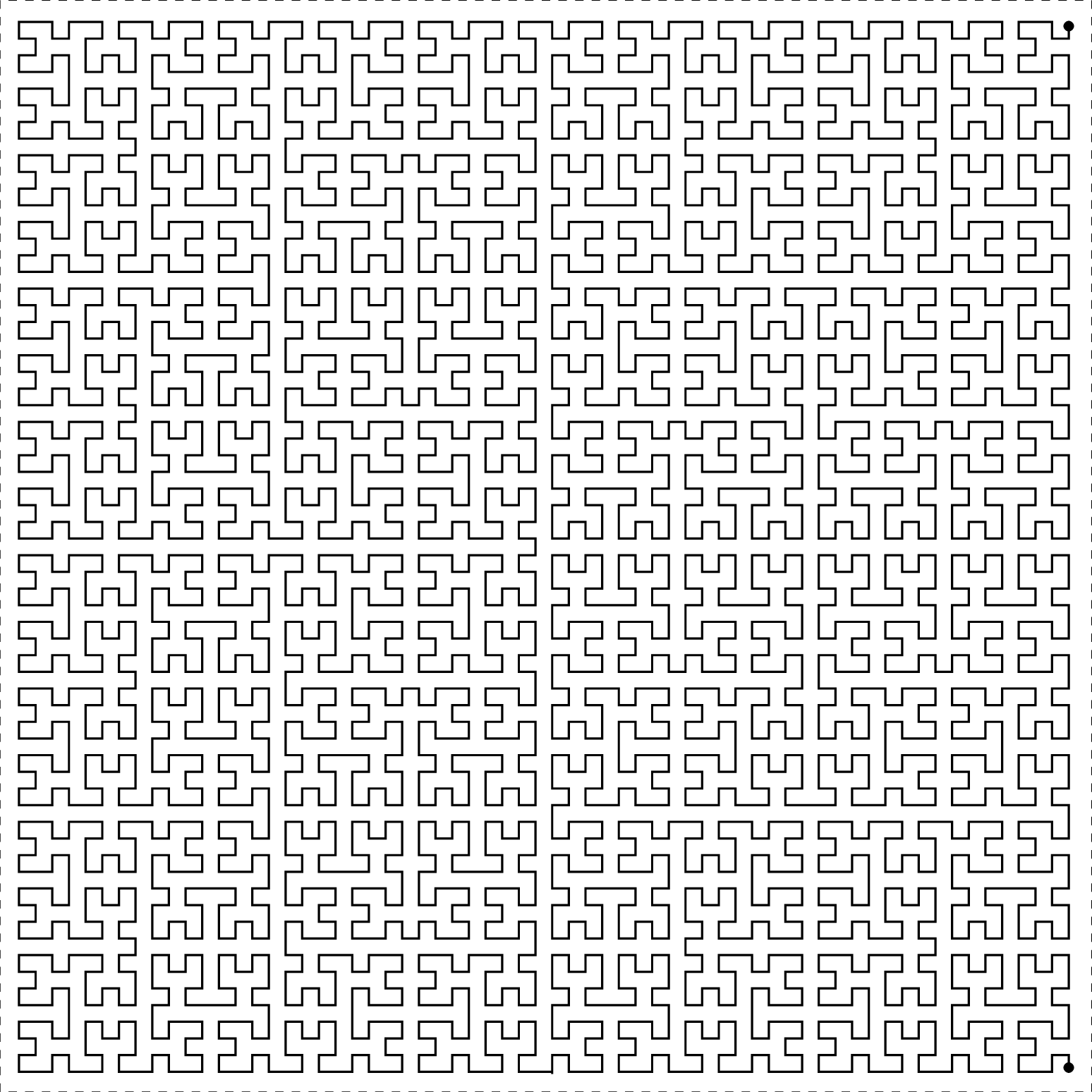}
\caption{Hilbert space-filling curve, level 6}
\label{fig-sfc-h6}
\end{center}
\end{figure}

\begin{figure}[!htb]
\begin{center}
\includegraphics[width=1.\hsize]{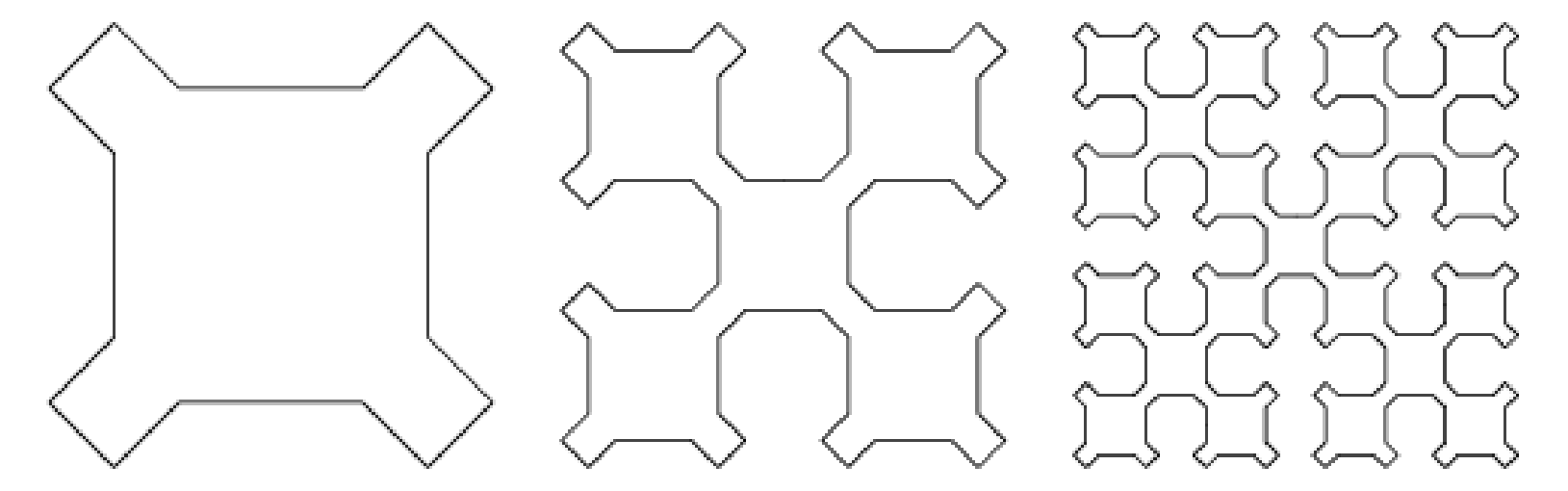}
\caption{Sierpi\'{n}ski space-filling curves, levels 1, 2 and 3}
    \label{fig-sfc-sie}
\end{center}
\end{figure}

\begin{figure}[!htb]
\begin{center}
\includegraphics[width=0.56 \hsize]{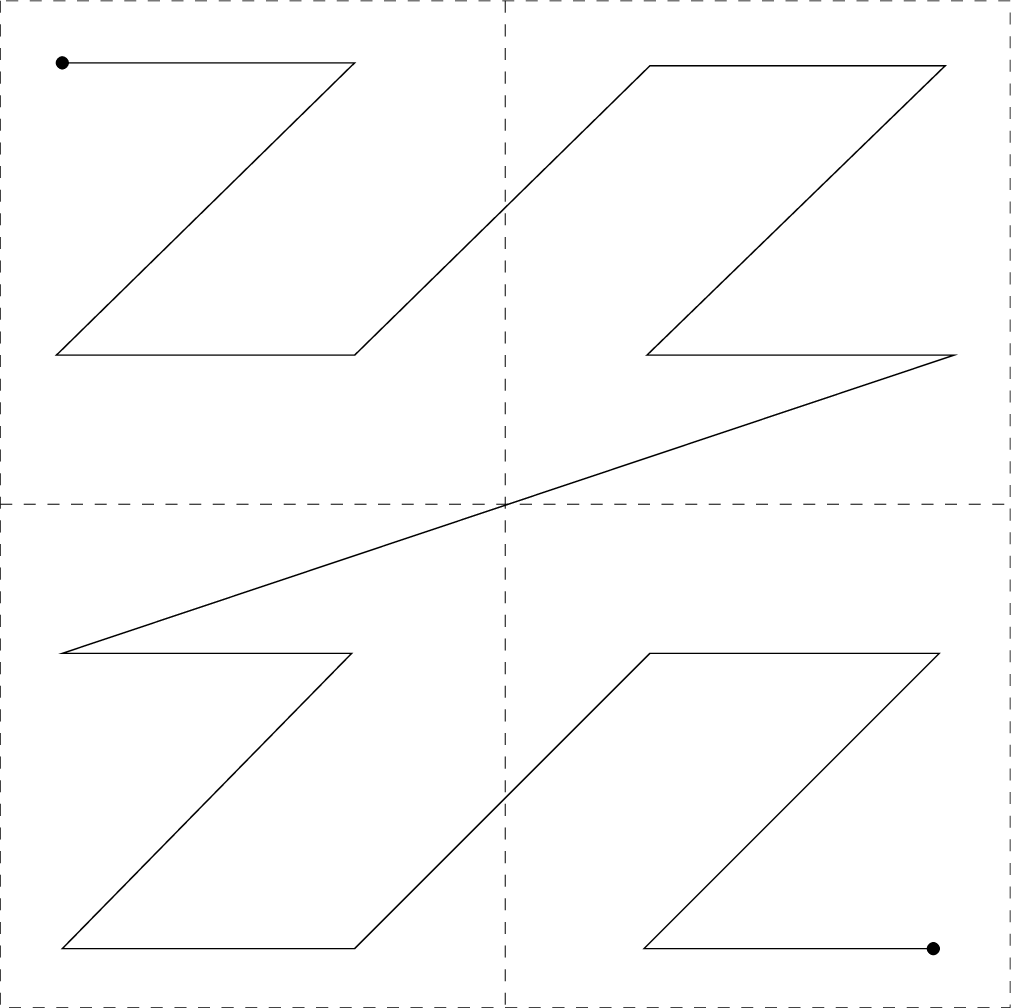}
\caption{Morton space-filling curve, level 2}
    \label{fig-sfc-z}
\end{center}
\end{figure}

Many space-filling curves have been discovered.
Figures \ref{fig-hilbert} and \ref{fig-sfc-h6} show levels 1, 2, 3 and 6 Hilbert space-filling curves
in a two-dimensional unit square. We can see that a curve is denser if a level is higher.
Figure \ref{fig-sfc-sie} shows levels 1, 2 and 3 Sierpi\'{n}ski space-filling curves.
Figure \ref{fig-sfc-z} shows a level 2 Morton space-filling curve.
From these curves, we can observe that the Hilbert space-filling curves and the Sierpi\'{n}ski space-filling curves
have good locality, and the Morton space-filling curves have jumps, whose locality is poor.

\subsubsection{Hilbert Orders}

\begin{figure}[!htb]
\begin{center}
\includegraphics[width=0.5 \hsize]{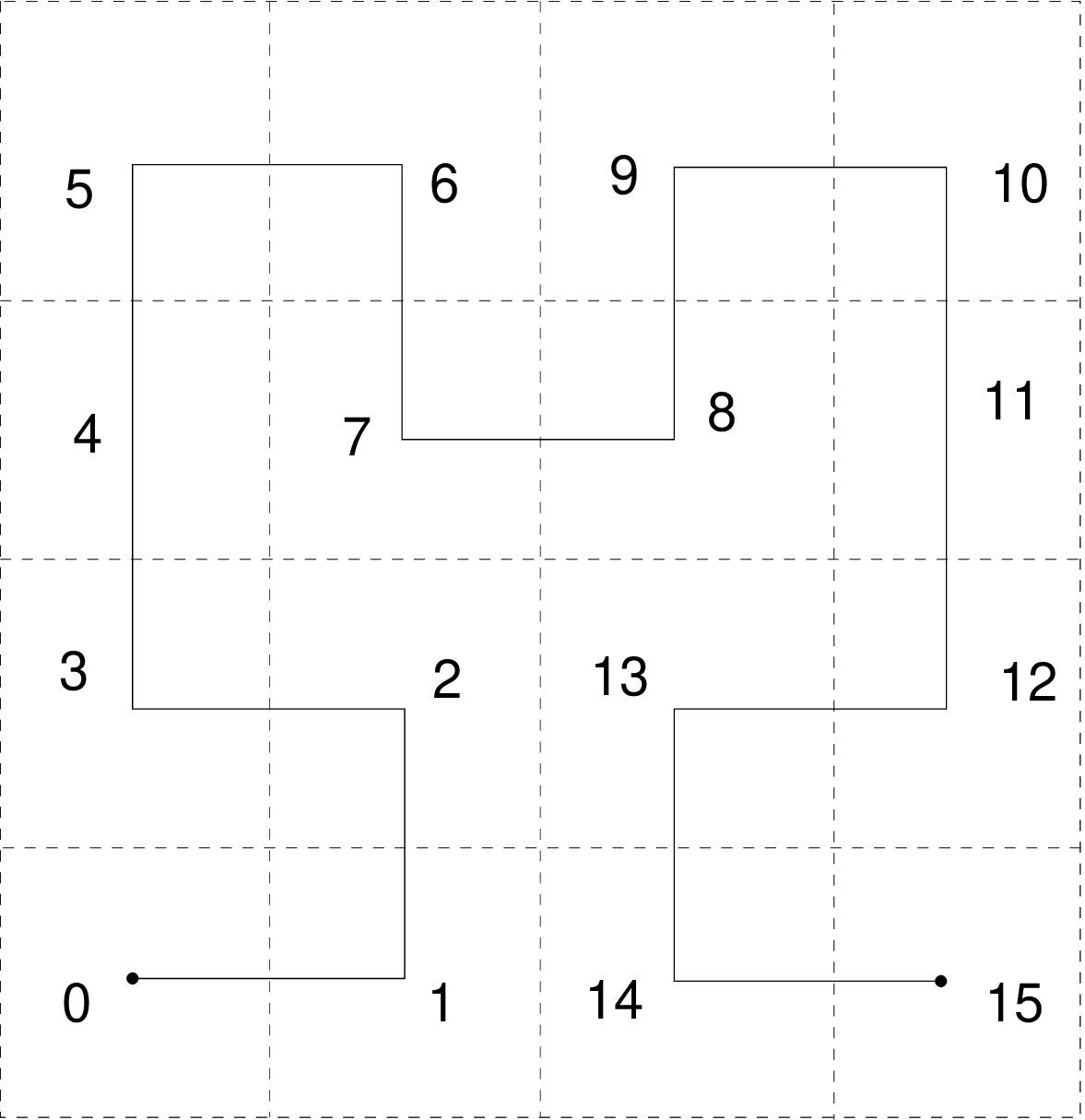}
\caption{Hilbert order, level 2} \label{fig-sfc-h2o}
\end{center}
\end{figure}

\begin{figure}[!htb]
\begin{center}
\includegraphics[width=0.5 \hsize]{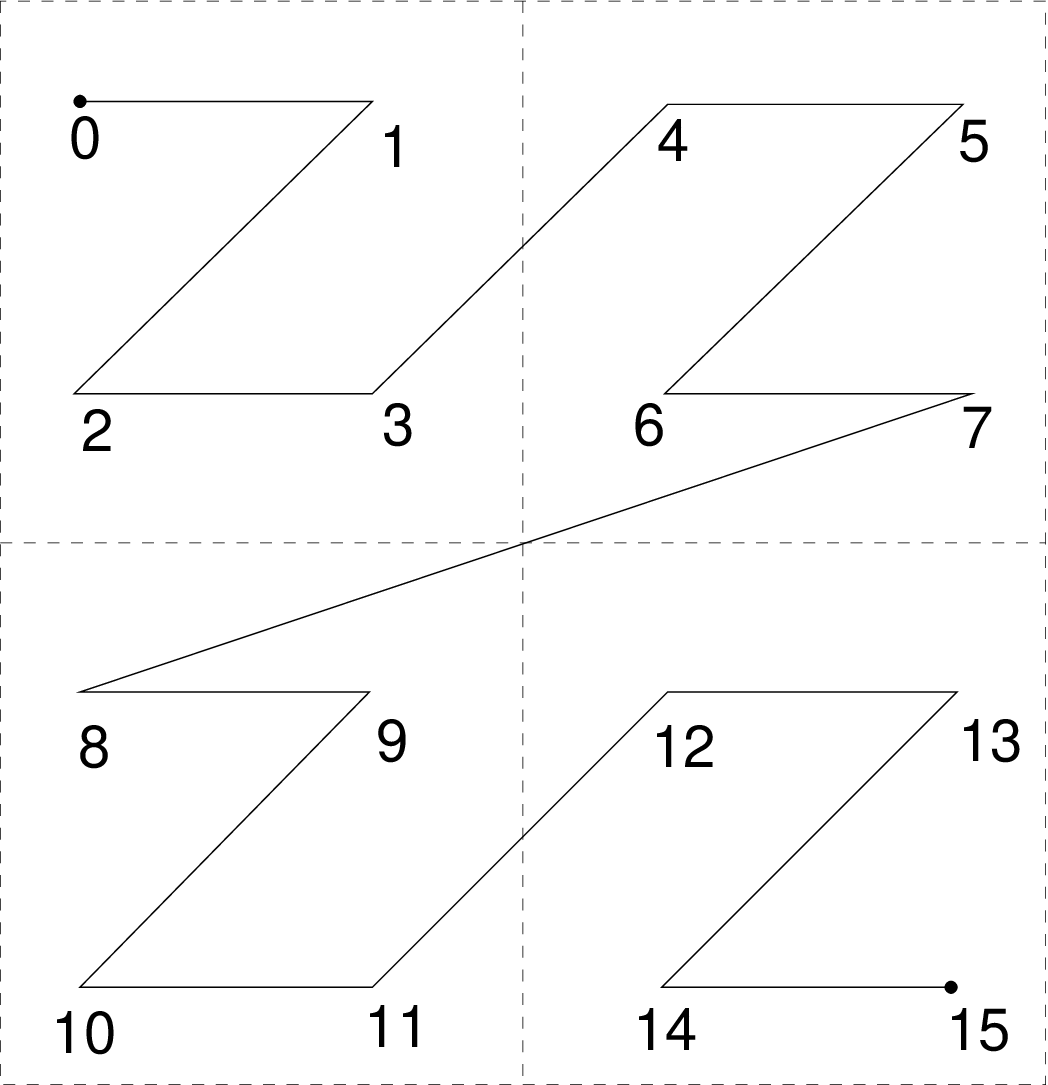}
    \caption{Morton order (Z-order), level 2} \label{fig-sfc-zo}
\end{center}
\end{figure}

Each curve has a starting point and an ending point. Along this curve,
a map is introduced between a one-dimensional domain and a multi-dimensional domain.
Figure \ref{fig-sfc-h2o} and \ref{fig-sfc-zo} show a level 2 Hilbert space-filling curve and a level 2 
Morton space-fill curve (Hilbert order), respectively. 
Both of them have 16 vertices whose indexes start from 0 to 15.
We can see that these curves define orders, which map a two-dimensional space to a one-dimensional space.
Higher dimensional space-filling curves are defined similarly.

A Hilbert space-filling curve is one famous space-filling curve, which is also called a Hilbert
curve \cite{Hans}. This kind of curve has many important characteristics, such as locality,
clustering and self-similarity. A Hilbert curve (order) has been applied in many areas,
including image storing, database indexing, data compression and dynamic load balancing.
For parallel computing, the Hilbert order method is one of the most important
geometry-based partitioning methods.

An $n ~ (n \ge 2)$-dimensional Hilbert curve introduces a one-to-one mapping between an
$n$-dimensional space and a one-dimensional space.
The mapping from the $n$-dimensional space to the one-dimensional space is called encoding,
while the inverse mapping is called decoding, which maps an integer to a coordinate in the $n$-dimensional space.
Algorithms for computing Hilbert curves/orders in two- and three-dimensional spaces have been proposed in
the literature, which can be classified into recursive algorithms
\cite{butz, gold, witten, cole} and iterative algorithms \cite{griff, sfc-tr-03, xliu2, xliu, fisher, ningtao}.
Iterative algorithms, especially the table-driven algorithms \cite{griff,
sfc-tr-03}, are usually much faster than recursive algorithms.
In general, the complexities of these algorithms are $O(m)$, where $m$ is
the level of a Hilbert curve. For a two-dimensional space,
Chen et al. \cite{ningtao} proposed an algorithm of $O(r)$ complexity,
where $r$ is defined as $r=\log_2(\max(x,y))+1$, $r \le m$ and is independent of the level $m$.
This algorithm is faster when $m$ is much larger than $r$.
The same idea was also applied to a three-dimensional space \cite{HSFC}.
For higher dimensional spaces, Kamata {et al.} presented a representative $n$-dimensional Hilbert
mapping algorithm \cite{kamata} and Li {et al.} introduced algorithms for analyzing the properties of
$n$-dimensional Hilbert curves \cite{chenyang}. Liu et al. introduced high-order encoding
and decoding algorithms \cite{liuh}.

\subsubsection{Calculation of Arbitrary Dimensional Hilbert Orders}

For the sake of completeness, calculations of Hilbert orders are introduced in this paper \cite{liuh}.
Let $m$ be the level of the Hilbert curves (orders) and $D_m$ be the coordinate set of the $m$th level Hilbert
curve, where $D_{m}= \{(x_n,\cdots, x_2, x_1)| 0 \le x_i < 2^{m}, 1
\le i \le n\}$. $(x_n, \cdots, x_1) ~ (\in D_m)$ is a coordinate of the Hilbert curve, and $x_i ~ ( 1 \le i \le n)$ is
called the $i$th component of the coordinate. A logical operation, \textbf{exclusive or operation (xor)},
$\wedge$ for two coordinates is defined as
\begin{equation}
(x_n, \cdots, x_2, x_1)\wedge (y_n, \cdots, y_2, y_1) = (x_n\wedge
y_n, \cdots, x_2\wedge y_2, x_1\wedge y_1).
\end{equation}
Here we use $(a_1a_2 \cdots a_k)_{d}$ to represent a number, where $ 0 \le
a_i < d ~ (1 \le i \le k)$, and $k$ can be any positive integer.
This number is a binary number if we set $d=2$ and a decimal number if we set $d=10$.
The \textbf{and} operator is denoted by $\&$. Let us define
\begin{equation}
p_n^{i}(a_1,a_2, \cdots, a_n) = (\sum\limits_{j = 1}^i {a_j}) ~mod~ 2 = (\sum\limits_{j = 1}^i {a_j}) ~\%~ 2,
\end{equation}
where $ a_i$ equals 0 or 1 $(1 \le i \le n)$. The $p_n^{i}(a_1,a_2, \cdots, a_n)$ equals 1 or 0.
With the help of $p_i$, function $f_n$ is defined as
\begin{equation}
\label{forward}
    f_n(a_1,a_2, \cdots ,a_n) = (b_1b_2\cdots b_n)_2 =
    j, b_1 = a_1, b_i =\left\{\begin{array}{rl}
        a_i  ~ if ~p_n^{i-1} = 0  \\
        1 - a_i ~ if ~p_n^{i - 1} = 1
\end{array}\right. ,
\end{equation}
where $a_i$ equals 0 or 1 and $j$ is a decimal number. This
function $f_n$ maps a vector $(a_1, a_2, \cdots, a_n)$ to a decimal number $j$.
Its inverse function $b_n$ is defined as
\begin{equation}
\label{backward} b_n(j) = b_n((a_1a_2 \cdots a_n)_2) =
(b_1,b_2,\cdots,b_n), b_1 = a_1, b_i =\left\{\begin{array}{rl}
a_i ~ if ~a_{i-1} = 0  \\
1 - a_i ~ if ~a_{i - 1} = 1
\end{array}\right. ,
\end{equation}
where $j ~ (0 \le j < 2^n)$ is a decimal number and $j = (a_1a_2 \cdots a_n)_2$.
This function maps a decimal number (scalar) to a vector.

For any point $x$, it can be mapped to an integer vector such that
$(x_n, \cdots, x_1) \in D_m$. Each component $x_i  ~ (1 \le i \le n)$
is written as $x_i = (x_i^mx_i^{m-1}\cdots x_i^1)_2$. The calculated Hilbert order is
stored as $(r_mr_{m-1}\cdots r_1)_{2^n}$, which will be mapped to $[0, 1]$ in the end.
The mapping is described in Algorithm \ref{alg-encoding1}, which is an iterative method.
Here we assume that $G_n^{i,0}$ and $G_n^{i,1}$ are known,
which can be calculated by the algorithms introduced in \cite{chenyang}.
This algorithm is for arbitrary dimensional cases. For a specific dimension, such as two dimensions,
special optimization techniques can be applied to accelerate calculations \cite{liuh}.

\begin{algorithm}[!htb]
\caption{Hilbert order algorithm}
\label{alg-encoding1}
\begin{algorithmic}
\State (1) For $ x \in [0, 1]^n$, map $x$ to domain $D_m$ by multiplying $2^m$. Let $k = m$.
\State (2) If $k = 0$, terminate the procedure. Or we have $r_k =
f_n(x_n^k, x_{n-1}^k, \cdots, x_1^k)$.

\State (3) For each integer $i ~ (1 \le i \le n)$, if $x_i^k$ equals 1, then $x_i = x_i - 2^{k-1}$.

\State (4) For each integer $i ~ (1 \le i \le n)$, if the $i$th component of $G_n^{{r_k},1}$ is 1, then $x_i = 2^{k-1}
- 1 - x_i$.

\State (5) If $G_n^{{r_k},0}$ has two components equal to 1 in the $i$th and $j$th positions, then swap $x_i$ and
$x_j$.

\State (6) $k = k - 1$, go to (1).
\State (7) $h = \frac{(r_mr_{m-1}\cdots r_1)_{2^n}}{{2^n}^m}$ ($h \in [0, 1]$).
\end{algorithmic}
\end{algorithm}

\subsubsection{Table-driven Algorithms for Low Dimensional Hilbert Orders}

Table-driven algorithms were introduced in \cite{griff, sfc-tr-03}. The basic idea is
to store additional information other than to calculate. More memory
is required, but computations are faster.

\begin{figure}[!htb]
\begin{evb}
typedef double FLOAT;
typedef int    INT;
typedef int    BOOLEAN;

typedef struct HSFC_ENTRY_ {
    FLOAT coord[3];
    FLOAT hsfc;

} HSFC_ENTRY;

static INT hsfc_maxlevel  = 30;
\end{evb}
    \caption{Data types for Hilbert order}
    \label{ho-ds}
\end{figure}

\begin{figure}[!htb]
\begin{evb}
static unsigned const int idata3d[] = {
    0, 7, 3, 4, 1, 6, 2, 5, 0, 1, 3, 2, 7, 6, 4, 5,
    0, 3, 7, 4, 1, 2, 6, 5, 2, 3, 5, 4, 1, 0, 6, 7,
    4, 5, 3, 2, 7, 6, 0, 1, 4, 7, 3, 0, 5, 6, 2, 1,
    6, 7, 5, 4, 1, 0, 2, 3, 0, 1, 7, 6, 3, 2, 4, 5,
    2, 1, 5, 6, 3, 0, 4, 7, 6, 1, 5, 2, 7, 0, 4, 3,
    0, 7, 1, 6, 3, 4, 2, 5, 2, 1, 3, 0, 5, 6, 4, 7,
    4, 7, 5, 6, 3, 0, 2, 1, 4, 5, 7, 6, 3, 2, 0, 1,
    6, 1, 7, 0, 5, 2, 4, 3, 0, 3, 1, 2, 7, 4, 6, 5,
    2, 3, 1, 0, 5, 4, 6, 7, 6, 7, 1, 0, 5, 4, 2, 3,
    2, 5, 1, 6, 3, 4, 0, 7, 4, 3, 7, 0, 5, 2, 6, 1,
    4, 3, 5, 2, 7, 0, 6, 1, 6, 5, 1, 2, 7, 4, 0, 3,
    2, 5, 3, 4, 1, 6, 0, 7, 6, 5, 7, 4, 1, 2, 0, 3
};
\end{evb}
    \caption{Ordering table for Hilbert order}
    \label{ho-odtable}
\end{figure}

\begin{figure}[!htb]
\begin{evb}
static unsigned const int istate3d[] = {
    1,  6,  3,  4,  2,  5,  0,  0,  0,  7,  8,  1,  9,  4,  5,  1,
    15, 22, 23, 20, 0,  2,  19, 2,  3,  23, 3,  15, 6,  20, 16, 22,
    11, 4,  12, 4,  20, 1,  22, 13, 22, 12, 20, 11, 5,  0,  5,  19,
    17, 0,  6,  21, 3,  9,  6,  2,  10, 1,  14, 13, 11, 7,  12, 7,
    8,  9,  8,  18, 14, 12, 10, 11, 21, 8,  9,  9,  1,  6,  17, 7,
    7,  17, 15, 12, 16, 13, 10, 10, 11, 14, 9,  5,  11, 22, 0,  8,
    18, 5,  12, 10, 19, 8,  12, 20, 8,  13, 19, 7,  5,  13, 18, 4,
    23, 11, 7,  17, 14, 14, 6,  1,  2,  18, 10, 15, 21, 19, 20, 15,
    16, 21, 17, 19, 16, 2,  3,  18, 6,  10, 16, 14, 17, 23, 17, 15,
    18, 18, 21, 8,  17, 7,  13, 16, 3,  4,  13, 16, 19, 19, 2,  5,
    16, 13, 20, 20, 4,  3,  15, 12, 9,  21, 18, 21, 15, 14, 23, 10,
    22, 22, 6,  1,  23, 11, 4,  3,  14, 23, 2,  9,  22, 23, 21, 0
};
\end{evb}
    \caption{Orientation table for Hilbert order}
    \label{ho-ottable}
\end{figure}

\begin{figure}[!htb]
\begin{evb}
    static unsigned const int *d[] = {
        idata3d,       idata3d + 8,   idata3d + 16,  idata3d + 24,
        idata3d + 32,  idata3d + 40,  idata3d + 48,  idata3d + 56,
        idata3d + 64,  idata3d + 72,  idata3d + 80,  idata3d + 88,
        idata3d + 96,  idata3d + 104, idata3d + 112, idata3d + 120,
        idata3d + 128, idata3d + 136, idata3d + 144, idata3d + 152,
        idata3d + 160, idata3d + 168, idata3d + 176, idata3d + 184
    };

    static unsigned const int *s[] = {
        istate3d,       istate3d + 8,   istate3d + 16,  istate3d + 24,
        istate3d + 32,  istate3d + 40,  istate3d + 48,  istate3d + 56,
        istate3d + 64,  istate3d + 72,  istate3d + 80,  istate3d + 88,
        istate3d + 96,  istate3d + 104, istate3d + 112, istate3d + 120,
        istate3d + 128, istate3d + 136, istate3d + 144, istate3d + 152,
        istate3d + 160, istate3d + 168, istate3d + 176, istate3d + 184
    };
\end{evb}
    \caption{State tables for Hilbert order}
    \label{ho-sttable}
\end{figure}

\begin{figure}[!htb]
\begin{evb}
void HilbertInvOrder3d(HSFC_ENTRY *x)
{
    int level, EffLen;
    unsigned int key[3], c[3], temp, stat;
    INT i;

    static unsigned INTMX;
    static unsigned EfBit;
    static BOOLEAN initialized = FALSE;
    static int k0 = 0, k1 = 0, k2 = 0;

    if (!initialized) {
        initialized = TRUE;

        INTMX = 4294967295U;
        EfBit = INTMX >> 2;

        k0 = 60 - hsfc_maxlevel * 3;
        k1 = 30 - hsfc_maxlevel * 3;
        k2 = -hsfc_maxlevel * 3;
    }

    c[0] = (unsigned int)(x[i].coord[0] * (double)INTMX);
    c[1] = (unsigned int)(x[i].coord[1] * (double)INTMX);
    c[2] = (unsigned int)(x[i].coord[2] * (double)INTMX);
    c[1] >>= 1;
    c[2] >>= 2;

    key[0] = key[1] = key[2] = 0;
    stat = 0;
    EffLen = 30;
    for (level = 0; level < hsfc_maxlevel; level++) {
        EffLen--;
        temp = ((c[0] >> EffLen) & 4) | ((c[1] >> EffLen) & 2) | ((c[2] >> EffLen) & 1);

        key[0] = (key[0] << 3) | ((key[1] >> 27) & 7);
        key[1] = (key[1] << 3) | ((key[2] >> 27) & 7);
        key[2] = (key[2] << 3) | *(d[stat] + temp);

        stat = *(s[stat] + temp);
    }

    key[0] = key[0] & EfBit;
    key[1] = key[1] & EfBit;
    key[2] = key[2] & EfBit;

    x[i].hsfc  = ldexp((double)key[2], k2);
    x[i].hsfc += ldexp((double)key[1], k1);
    x[i].hsfc += ldexp((double)key[0], k0);
}
\end{evb}
    \caption{Computation of Hilbert order}
    \label{ho-algtable}
\end{figure}

In \cite{sfc-tr-03}, the authors introduced the Gray coding and Morton ordering, which are
easy to compute. Other orderings, such as the
Hilbert ordering, can be generated through pre-defined appropriate mappings. An
ordering table and an orientation table are required to map between Hilbert and Morton orders.
Figure \ref{ho-ds} shows basic data structures and data types, from which integers, floating-point
numbers and Hilbert orders are defined. Figures \ref{ho-odtable} and \ref{ho-ottable} are
the ordering table and orientation table between the Morton and Hilbert orders.
Figure \ref{ho-sttable} shows internal state conversion rules.
Figure \ref{ho-algtable} is a C code for generating a Hilbert order in a three-dimensional
space, which maps $[0, 1]^3$ to $[0, 1]$.

\subsubsection{Space-filling Curve Partitioning Method}

Space-filling curve methods are good alternatives for the graph methods,
which can provide good partitioning. These
methods are faster compared with the graph methods. Algorithm \ref{alg-hsfc} shows the
process of the space-filling curve methods, which has three steps.
The first step is to map a computational domain $\Omega$ to a subset of $(0, 1)^3$, which
can be obtained by a linear mapping. Then, for any cell, its new centroid coordinates belong to
$(0, 1)^3$. A space-filling curve defines a map from $[0, 1]^3$ to $[0, 1]$.
We can employ this map to calculate the values of the new centroid coordinates in $(0, 1)$.
The third step is to partition $(0, 1)$ into $N_p$ sub-intervals such that each sub-interval has
the same numbers of cells (or workload). Each MPI can perform the first and the second steps
independently. However, in our implementation, the third step is calcuated by one MPI and it
broadcasts results to other MPIs.

These methods assume that two cells (elements) that are close to each other
have a higher possibility to communicate with each other than two cells that are far from each other.
This assumption is true for grid-based numerical methods,
such as the finite element, finite volume and finite difference methods.
The only difference for various space-filling curve methods is how to map a cell to $(0, 1)$.

\begin{algorithm}[!htb]
\caption{Space-filling curve method}
\label{alg-hsfc}
\begin{algorithmic}[1]
\State Map a computational domain $\Omega$ to a subset of $(0, 1)^3$.
\State For any cell, calculate its mapping that belongs to $(0, 1)$.
\State Partition the interval $(0, 1)$ into $N_p$ sub-intervals and each sub-interval has the same number of cells.
\end{algorithmic}
\end{algorithm}

\subsubsection{Partition Quality}

This section studies partition quality on homogeneous architectures.
The quality of a partition resulting from the graph methods and space-filling curve methods
has several measurements, including a load imbalance factor, a local surface index
and a global surface index. These concepts have been studied in \cite{liuh}.

The workload distribution is measured by the load imbalance factor.
It is reasonable to assume that all cells have equal computations
so the workload of each processor can be approximated by the number of cells it owns.
The load imbalance factor $f_p$ is defined as follows:
\begin{equation}
f_p = \frac{N_p \times \max_{1 \le i \leq N_p} |\mathbb{G}_i|} {\sum_{i = 1}^{N_p} |\mathbb{G}_i|}.
\end{equation}

Let $f_i$ be the number of faces in the sub-grid $\mathbb{G}_i|$ and $b_i$ be the number of faces
that are shared by a cell in another processor.
We should mention that $b_i$ is proportional to the communication volume of the $i$th processor involved.

The maximum local surface index is defined by
\begin{equation}
r_M = \max_{0\le i < n}\frac{b_i}{f_i}.
\end{equation}
It is used to model the maximal communications that one processor involves.
The global surface index is defined by
\begin{equation}
r_G = \frac{\sum_{i = 1}^{N_p}{b_i}}{\sum_{i = 1}^{N_p}{f_i} - \sum_{i = 1}^{N_p}{b_i}}.
\end{equation}
It models the overall communications of all processors.
The average surface index is defined by
\begin{equation}
    r_M = \sum_{i = 1}^{N_p}\frac{b_i}{f_i}.
\end{equation}
It is used to model the average communications that one processor involves.

If a cell in a sub-grid has a neighbour in another processor, then we can say that these two
sub-grids are connected. Inter-processor connectivity is the number of connected sub-grids of
a given sub-grid, denoted by $c_i$. The maximal inter-processor connectivity is defined by
\begin{equation}
c = {\max_{1 \le i \leq N_p} c_i}.
\end{equation}

\section{DOF (Degrees of Freedom)}

The cell-centered data is natural to reservoir simulation, since each cell can represent
a block of a real oil and gas field and we can attach properties to the cell.
Its data structure, \verb|DOF|, is shown in Figure \ref{fig-ds-dof}.

\begin{figure}[!htb]
\centering
\begin{evb}
typedef struct DOF_TYPE_
{
    INT           np_cell;        /* number of DOFs per cell */
    INT           np_well;        /* number of DOFs per well */

} DOF_TYPE;

typedef struct DOF_
{
    char          *name;        /* name of DOF */
    GRID          *g;           /* the grid */
    DOF_TYPE      *type;        /* type of DOF */
    FLOAT         *data;
    INT           *idata;
    INT           count_cell;   /* data count per cell */
    INT           count_well;   /* data count per well */
    INT           count_perf;   /* data count per perforation */
    INT           dim;

    BOOLEAN       assembled;

} DOF;
\end{evb}
\caption{Data structure of DOF}
\label{fig-ds-dof}
\end{figure}

Each \verb|DOF| has a name (\verb|name|), and it associates with a grid (\verb|g|). 
It also has a type (\verb|type|).
It can define floating-point number and integer data. A \verb|DOF| also has a dimension (\verb|dim|),
through which scalar and vector can be defined.
In reservoir simulations, cells have different properties, such as oil and water saturations,
porosity and permeability.

\begin{figure}[!htb]
\centering
\begin{evb}
DOF_TYPE DOF_CELL_      = {1, 0};
DOF_TYPE DOF_WELL_      = {0, 1};
DOF_TYPE DOF_PERF_      = {0, 1};
DOF_TYPE DOF_CONSTANT_  = {1, 0};

DOF_TYPE *DOF_CELL     = &DOF_CELL_;
DOF_TYPE *DOF_WELL     = &DOF_WELL_;
DOF_TYPE *DOF_PERF     = &DOF_PERF_;
DOF_TYPE *DOF_CONSTANT = &DOF_CONSTANT_;
\end{evb}
\caption{DOF types}
\label{fig-ds-dof-type}
\end{figure}

DOF types are defined in Figure \ref{fig-ds-dof-type}.
\verb|DOF_TYPE_CELL| defines data on each cell.
\verb|DOF_TYPE_WELL| defines data on each well.
One well may perforate several cells and some properties only
exist on these perforated cells, such as oil production rate and water injection rate. 
We have another type, \verb|DOF_TYPE_PERF|, which has value on all perforated cells only.
\verb|DOF_CONSTANT| is for constant. The data types are floating-point number.

\begin{figure}[!htb]
\centering
\begin{evb}
typedef struct DOF_NBR_
{
    DOF      *dof;
    FLOAT    *data;
    INT      *idata;
    INT      dim;
    BOOLEAN  assembled;

} DOF_NBR;
\end{evb}
\caption{Remote DOF data}
\label{fig-ds-dof-remote}
\end{figure}

Cells may be in different MPIs, and sometimes, a cell requires data in its remote neighbouring
cell. \verb|DOF_NBR| is defined to gather remote DOF data for cell's direct neighbours.

Input and output (reading data from a file and writing data to a file) for sequential applications
are trivial, and
simple writing and reading functions from operating systems or C language are enough. However,
when we are working on parallel computing, each MPI task has portion of the whole grid and portion
of data, and each task only reads and writes part of data.
If each MPI task reads the whole input file and picks necessary data, then most read data will be dropped and they
compete disk with each other. If all MPI tasks write data to the same file, conflicts may exist.
The platform provides parallel input and output modules using MPI-IO,
which supports integer and floating-point numbers.

The initialization of reservoir simulations requires read reservoir properties from files, such as porosity and permeability.
The restart of simulations also needs to write intermediate data and to read data files.
They are achieved by our parallel input and output modules.

\section{MAP}

\begin{figure}[!htb]
\centering
\begin{evb}
typedef struct COMM_INFO_
{
    INT       *sidx;
    INT       *widx;
    int       *scnts, *sdsps;
    int       *rcnts, *rdsps;
    MPI_Comm  comm;
    INT       ssize, rsize;

} COMM_INFO;
\end{evb}
\caption{Data structure of communication info}
\label{fig-ds-comm}
\end{figure}

Figure \ref{fig-ds-comm} shows communication info used by matrix and vector. This data structure can be used
for collection communications, such as \verb|MPI_Alltoallv|, and point-to-point communications.
\verb|scnts| means data amounts sent by current MPI task to other MPI tasks, and \verb|sdsps| means the
displacement relative to sending buffer. \verb|rcnts| means data amounts received from other MPI tasks, and
\verb|rdsps| is the displacement relative to receiving buffer. \verb|ssize| is total data sent by current
MPI task and \verb|rsize| is total received data. \verb|sidx| is the index of data to be sent.
\verb|widx| is the local index of data received in current MPI task.

\begin{figure}[!htb]
\centering
\begin{evb}
typedef struct LG_MAP_
{
    INT *Gmap;
    INT *Lmap;
    INT size;

} LG_MAP;

typedef struct MAP_
{
    DOF          **_dofs;
    INT          ndofs;

    LG_MAP       *_RNGmap;
    INT          num_rnmaps;

    INT          *_L2Gmap;
    INT          *part;
    INT          *offset;

    INT          nlocal;
    INT          ntlocal;
    INT          nglobal;
    int          refcount;

    LG_MAP       *_W2V;
    INT          *_W2L;
    INT          num_w2vs;
    LG_MAP       **_P2V;
    INT          *num_perfs;
    INT          nwells_global;

    MPI_Comm     comm;
    int          rank, nprocs;
    COMM_INFO    *cinfo;

    BOOLEAN      assembled;

} MAP;
\end{evb}
\caption{Data structure for MAP}
\label{fig-ds-map}
\end{figure}

A map, \verb|MAP|, is defined to store communication information among cell data, matrices and vectors. It includes \verb|DOF| information,
off-process entries, communication pattern, locations of data required by other MPI tasks and locations of data
received from other MPI tasks.

\section{Distributed Matrices and Vectors}
A linear system, $Ax = b ~ (A \in R^{N \times N})$, is assembled in each Newton iteration.
Distributed matrix and vector are required to store the linear system. In the platform, each matrix
and vector are distributed among all MPI tasks.

Each row of the distributed matrix has a unique global
row index, which ranges from 0 to $N - 1$ consecutively and is numbered from the 1-th 
MPI task to the $N_p$-th MPI task.
Each row also has a local index on each MPI task.
The global indices of a vector is numbered the same way.

\subsection{Vector}
A distributed floating-point vector is defined in Figure \ref{fig-ds-vec},
which has buffer that holds data entries (\verb|data|),
number of local entries belong to current MPI task (\verb|nlocal|),
number of total entries (including off-process entries, \verb|ntlocal|) and reference to mapping information.

\begin{figure}[!htb]
\centering
\begin{evb}
typedef struct VEC_
{
    MAP      *map;
    FLOAT    *data;
    INT      nlocal;    /* entries belong to current proc */
    INT      ntlocal;   /* total entries in current MPI process */

} VEC;
\end{evb}
\caption{Data structure of VEC}
\label{fig-ds-vec}
\end{figure}

\subsection{Matrix}

\begin{figure}[!htb]
\centering
\begin{evb}
/* struct for a matrix row */
typedef struct MAT_ROW_
{
    FLOAT   *data;     /* data */
    INT     *cols;     /* local column indices, INT[ncols] */
    INT     *gcols;    /* global column indices, INT[ncols] */
    INT     ncols;     /* number of nonzero columns */

} MAT_ROW;

typedef struct MAT_
{
    MAT_ROW    *rows;

    MAP        *map;
    COMM_INFO  *cinfo;
    INT        *O2Gmap;

    INT        nlocal;       /* local entries belong to current proc */
    INT        ntlocal;      /* total local entries */
    INT        nglobal;      /* global matrix size */
    INT        *part;        /* distribution information */

    int        rank;
    int        nprocs;
    MPI_Comm   comm;
    BOOLEAN    assembled;

} MAT;
\end{evb}
\caption{Data structure of MAT}
\label{fig-ds-mat}
\end{figure}

The data structure of a distributed matrix is more complex than a vector, which requires entries for each row
and some other additional information. It is represented in Figure \ref{fig-ds-mat}. The \verb|MAT_ROW| stores
non-zero entries of each row and its storage format is similar to a CSR matrix,
which has a value of an entry (\verb|data|), the global index (\verb|gcol|)
and local index of an entry (\verb|col|). The \verb|MAT| has communication information (\verb|cinfo|),
MPI information and additional information, such as mapping between local index of off-process entries and their
global indices (\verb|O2Gmap|), and row distribution among all MPI tasks (\verb|part|).

\subsection{Algebraic Operations}

With the help of above data structures, commonly used matrix-vector operations and vector
operations can be implemented directly, which are listed as follows:
\begin{equation}
y = \alpha A x + \beta y,
\end{equation}
\begin{equation}
z = \alpha A x + \beta y,
\end{equation}
\begin{equation}
y = \alpha x + \beta y,
\end{equation}
\begin{equation}
z = \alpha x + \beta y,
\end{equation}
\begin{equation}
\alpha = \langle x, y \rangle,
\end{equation}
\begin{equation}
\alpha = \langle x, x \rangle^{\frac{1}{2}},
\end{equation}
where $A$ is a matrix, $\alpha$ and $\beta$ are scalars, and $x$ and $y$ are vectors.

\section{Linear System}
For the linear system, $A x = b$, derived from a nonlinear method, Krylov subspace solvers
including the restarted GMRES(m) solver, the BiCGSTAB solver,
and algebraic multi-grid (AMG) solvers are commonly used to find its solution.
The Krylov subspace solvers mentioned here are
suitable for arbitrary linear systems while the algebraic multi-grid solvers are efficient for positive
definite linear systems.

\subsection{Solvers}
In-house parallel Krylov subspace linear solvers are implemented.
The data structure of our solvers, \verb|SOLVER|, is shown in Figure \ref{fig-ds-solver}, which includes
parameters (\verb|rtol|, \verb|atol|, \verb|btol|, \verb|maxit|, \verb|restart|),
matrices, right-hand sides, solutions, and preconditioner information.

\begin{figure}[!htb]
\centering
\begin{evb}
typedef struct SOLVER_
{
    FLOAT                 residual;
    FLOAT                 rtol;
    FLOAT                 atol;
    FLOAT                 btol;
    int                   nits;
    INT                   maxit;
    INT                   restart;

    MAT                   *A;
    VEC                   *rhs;
    VEC                   *x;
    MAP                   *map;

    /* preconditioner */
    SOLVER_PC             pc;
    PC_TYPE               pc_type;

    int                   rank;
    int                   nprocs;
    MPI_Comm              comm;
    BOOLEAN               assembled;
    BOOLEAN               oem_created;

} SOLVER;

\end{evb}
\caption{Data structure of SOLVER}
\label{fig-ds-solver}
\end{figure}

\subsection{Preconditioners}

Several preconditioners are developed, including general purpose preconditioners and physics-based preconditioners
for reservoir simulations only.

\subsubsection{Restricted Additive Schwarz Method}

For the classical ILU methods,
the given matrix $A$ is factorized into a lower triangular matrix $L$ and an upper triangular matrix $U$;
a lower triangular linear system and an upper triangular linear system are required to solve:
\begin{equation}
L U x = b \Longleftrightarrow L y = b, U x = y.
\end{equation}
The systems need to be solved row-by-row, which are serial.
It is well-known that they have limited scalability.
Another option for parallel computing is the restricted additive Schwarz (RAS) method \cite{DDM},
which was developed by Cai et al.

\begin{figure}[!htb]
\centering
\begin{evb}
typedef struct RAS_PARS_
{
    INT         overlap;
    INT         iluk_level;
    INT         ilut_p;
    int         solver;
    FLOAT       ilut_tol;
    FLOAT       filter_tol;
    INT         ilutc_drop;

} RAS_PARS;

/* RAS */
typedef struct RAS_DATA_
{
    COMM_INFO            *cinfo;
    mat_csr_t            L;
    mat_csr_t            U;

    FLOAT                *frbuf;
    FLOAT                *fxbuf;
    FLOAT                *fbbuf;
    INT                  *ras_pro;     /* ras prolongation */
    INT                  num_ras_pro;

    RAS_PARS             pars;

} RAS_DATA;
\end{evb}
\caption{Data structure of RAS preconditioner}
\label{fig-ds-ras}
\end{figure}

\begin{figure}[!htb]
\centering
\begin{evb}
static RAS_PARS ras_pars_default =
{
    /* overlap */       1,
    /* k */             0,
    /* ilut_p */        -1,
    /* solver */        ILUK,
    /* ilut_tol */      1e-3,
    /* filter tol */    1e-4,
    /* drop */          0,
};
\end{evb}
\caption{Default parameters of RAS preconditioner}
\label{fig-ds-ras-par}
\end{figure}

The data structure of the RAS preconditioner is shown in Figure \ref{fig-ds-ras}. The \verb|pars|
stores parameters of the RAS preconditioner, such as overlap, local solver (ILUK, ILUT and ILTC),
the level of ILUK, memory control and tolerance of ILUT, and filter tolerance. The \verb|RAS_DATA|
has a local problem stored by the lower triangular matrix \verb|L| and the upper triangular matrix \verb|U|,
communication information of different sub-domains (\verb|cinfo|),
memory buffer and prolongation (restriction) operation information.

The default parameters of the RAS preconditioner is shown in Figure \ref{fig-ds-ras-par}.
Its default local solver is ILU(0). Default parameters for ILUT(p, tol) is -1 and 1e-3.
If $p$ is -1, the factorization subroutine will determine dynamically.

\subsubsection{Algebraic Multigrid Methods}

\begin{figure}[!htb]
    \centering
    \includegraphics[width=0.650\linewidth]{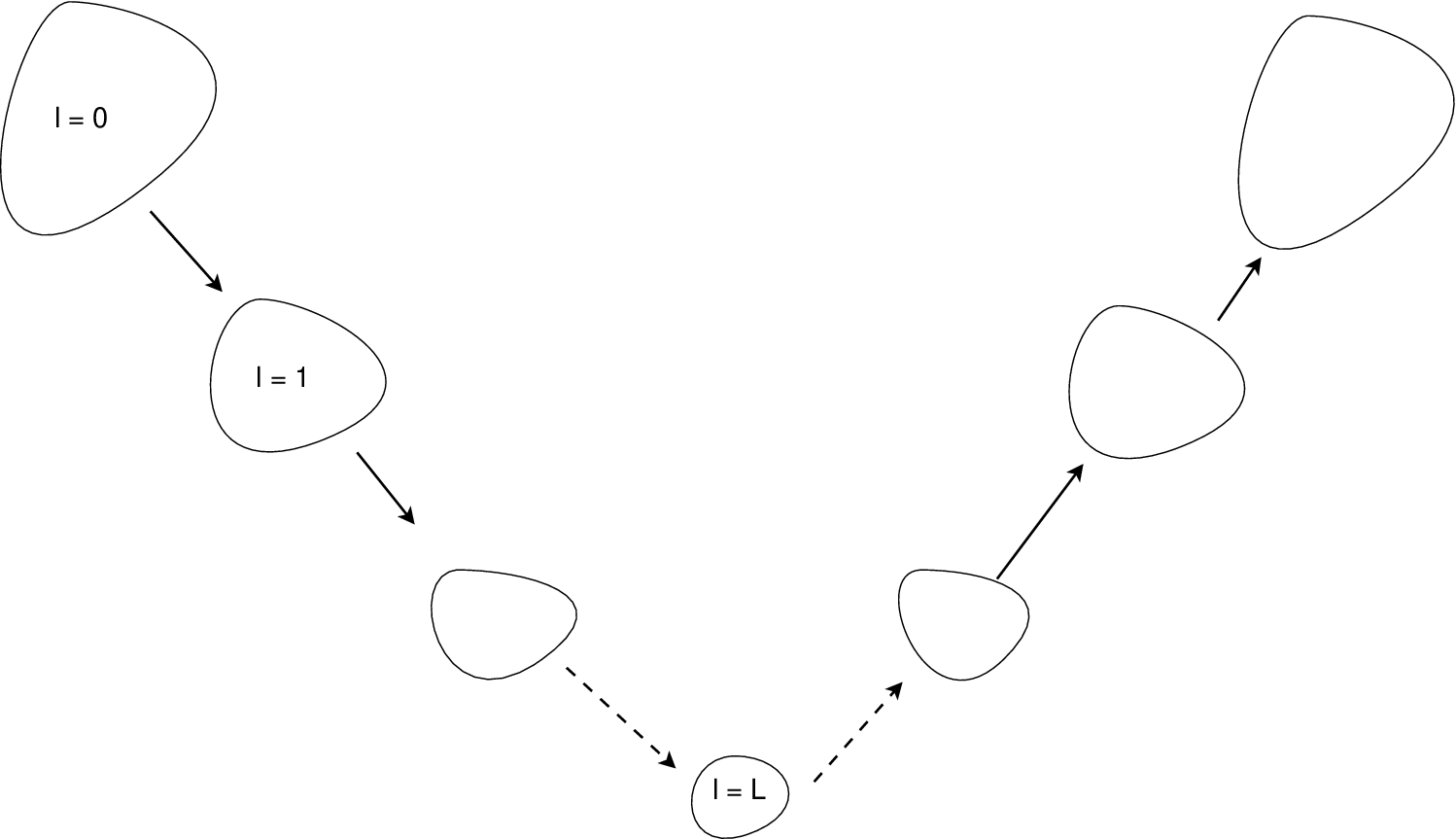}
    \caption{Structure of AMG solver.}
    \label{fig-amg}
\end{figure}

If $A$ is a positive-definite square matrix, the AMG methods
\cite{AMG, AMG-OR, Stuben2, BMR, falgout, psm} have proved to be
efficient methods and they are also scalable \cite{rsamg}.
AMG methods have hierarchical structures, and a coarse grid is chosen
when entering a coarser level.
Its structure of an algebraic multigrid solver is shown in Figure \ref{fig-amg}.

A restriction operator $R_l$ and
an interpolation (prolongation) operator $P_l$ are determined.
In general, the restriction operator $R_l$ is the transpose of the interpolation (prolongation) operator $P_l$:

\[
R_l = P_l^T.
\]
The matrix on the coarser grid is calculated by
\begin{equation}
A_{l + 1} = R_l A_l P_l.
\end{equation}
We know that a high frequency error is easier to converge on a fine grid than a low frequency error, and
for the AMG methods,
the restriction operator, $R_l$, projects the error from a finer grid onto a coarser grid and
converts a low frequency error to a high frequency error.
The interpolation operator transfers a solution on a coarser
grid to that on a finer grid.
Its setup phase for the AMG methods on each level $l$ ($0 \leq l < L$) is formulated in
Algorithm \ref{alg-amg-setup}, where a coarser grid, an interpolation operator, a restriction
operator, a coarser matrix and post- and pre-smoothers are constructed.
By repeating the algorithm, an $L$-level system can be built.
The solution of the AMG methods is recursive and is formulated in Algorithm \ref{alg-amg-solve}, which
shows one iteration of AMG. The AMG package we use is the BoomerAMG from HYPRE \cite{HYPRE2}.

The Cleary-Luby-Jones-Plassman (CLJP) parallel coarsening
algorithm was proposed by Cleary \cite{cljp1} based on the algorithms developed
by Luby \cite{cljp3} and Jones and Plassman \cite{cljp2}. The standard RS coarsening algorithm
has also been parallelized \cite{HYPRE2}. Falgout et al. developed a parallel coarsening algorithm,
the Falgout coarsening algorithm, which has been implemented in HYPRE \cite{HYPRE2}. Yang et al. proposed
HMIS and PMIS coarsening algorithms for a coarse grid selection \cite{pmis}. Various parallel smoothers
and interpolation operators have also been studied by Yang et al \cite{HYPRE2,pmis}.

\begin{algorithm}[!htb]
\caption{AMG setup Algorithm} \label{alg-amg-setup}
\begin{algorithmic}[1]
  \State Calculate strength matrix $S$.
  \State Choose coarsening nodes set $\omega_{l+1}$ according to strength matrix $S$, such that $\omega_{l+1} \subset \omega_{l}$.
  \State Calculate prolongation operator $P_l$.
  \State Derive restriction operator $R_l = P_l^T$.
  \State Calculate coarse matrix $A_{l+1}$: $A_{l+1} = R_l \times A_l \times P_l$.
  \State Setup pre-smoother $S_l$ and post-smoother $T_l$.
\end{algorithmic}
\end{algorithm}

\begin{algorithm}[!htb]
\caption{AMG V-cycle Solution Algorithm: amg\_solve($l$)} \label{alg-amg-solve}
\begin{algorithmic}%[1]
\State Require: $b_l$, $x_l$, $A_l$, $R_l$, $P_l$, $S_l$, $T_l$, $0 \leq l < L$
\State
\State $b_0 = b$

\If {($l < L$)}
  \State $x_l = S_l(x_l, A_l, b_l)$
  \State $r = b_l - A_lx_l$
  \State $b_{r+1} = R_lr$
  \State amg\_solve($l+1$)
  \State $x_l = x_l + P_l x_{l+1}$
  \State $x_l = T_l(x_l, A_l, b_l)$
\Else
  \State $x_l = A_l^{-1}b_l$
\EndIf

\State x = $x_0$

\end{algorithmic}
\end{algorithm}

\begin{figure}[!htb]
\centering
\begin{evb}
typedef struct BMAMG_PARS_
{
    INT         maxit;
    INT         num_funcs;           /* size of the system of PDEs */
    INT         max_levels;          /* max MG levels */
    FLOAT       strength;            /* strength threshold */
    FLOAT       max_row_sum;         /* max row sum */
    FLOAT       trunc_tol;           /* trunc tol */

    int         coarsen_type;        /* default coarsening type = Falgout */
    int         cycle_type;          /* MG cycle type */
    int         relax_type;          /* relaxation type */
    int         coarsest_relax_type; /* relax type on the coarsest grid */
    int         interp_type;         /* interpolation */
    INT         num_relax;           /* number of sweep */

} BMAMG_PARS;

typedef struct BMAMG_DATA_
{
    MAP                   *map;
    INT                   ilower, iupper;
    HYPRE_IJMatrix        A;
    HYPRE_IJVector        b, x;
    BOOLEAN               assembled;
    HYPRE_Solver          hsolver;

    SolveFcn              setup;
    SolveFcn              solve;
    DestroyFcn            destroy;

    BMAMG_PARS            pars;

} BMAMG_DATA;
\end{evb}
\caption{Data structure of AMG method}
\label{fig-ds-amg}
\end{figure}

\begin{figure}[!htb]
\centering
\begin{evb}
static BMAMG_PARS amg_pars_default =
{
    /* maxit */               1,
    /* num_funcs */           -1,
    /* max_levels */          6,
    /* strength */            0.5,
    /* max_row_sum */         0.9,
    /* trunc error */         1e-2,

    /* coarsen_type */        Falgout,
    /* cycle_type */          v-cycle,
    /* relax_type */          gs-h-forward,
    /* coarsest_relax_type */ gs-h-symmetric,
    /* interp type */         cmi,
    /* itr relax */           2,
};
\end{evb}
\caption{Default parameters of AMG method}
\label{fig-ds-amg-par}
\end{figure}

In our platform, the AMG solvers are from HYPRE. They can be used as solvers and preconditioners.
The data structure of the AMG solvers is shown in Figure \ref{fig-ds-amg}. The \verb|BMAMG_PARS| stores parameters
of the AMG method, including the coarsening type, interpolation type, maximal levels, smoother type,
and cycle type. The \verb|BMAMG_DATA| stores linear system information, such as a distribution pattern
of matrices and vectors, mapping information, and related interfaces.

Default parameters for the
AMG method is shown in Figure \ref{fig-ds-amg-par}, where a default six-level AMG method is applied.
The detailed explanation of each parameter can be read from the HYPRE user manual.

\subsubsection{CPR-like Preconditioners}

Linear systems from black oil, compositional and thermal models are hard to solve,
especially when the reservoirs are heterogeneous.
However, the matrices from the pressure unknowns are positive definite, which can be solved by
AMG methods. Many preconditioners have been developed to speed the solution of
linear systems, such as the constrained pressure residual (CPR) method and FASP method \cite{FASP}.
Here we introduce our multi-stage preconditioners for the black oil,
compositional and thermal models, which are based on the classical CPR method.

Numerical methods for black oil, thermal and compositional models may choose different unknowns \cite{Book-Chen}.
Here we assume that the oil phase pressure $p_o$ is always one of the unknowns.
The other variables are denoted as $\overrightarrow{s_i}$.
The well unknowns are denoted by $\overrightarrow{w}$, whose dimension equals
the number of wells in the reservoir, $N_w$. Let us define the pressure vector $p$ as:
\begin{equation}
p = \left(
        \begin{array}{c}
        p_{o,1} \\
        p_{o,2} \\
        \cdots   \\
        p_{o,N_g}
        \end{array}
        \right),
\end{equation}
and the global unknown vector $x$ as:
\begin{equation}
x = \left(
        \begin{array}{c}
         p_{o,1}   \\
        p_{o,2}    \\
         \cdots    \\
         p_{o,N_g} \vspace{0.2cm} \\
        \overrightarrow{s_1} \vspace{0.2cm}  \\
        \cdots   \\
        \overrightarrow{s_{N_g}} \vspace{0.2cm}  \\
        \overrightarrow{w}
        \end{array}
        \right).
\end{equation}
A restriction operator from $x$ to $p$ is defined as
\begin{equation}
\varPi_r  x = p.
\end{equation}
A prolongation operator $\varPi_p $ is defined as
\begin{equation}
\varPi_p  p = \left(
        \begin{array}{c}
        p \\
        \overrightarrow{0} \vspace{0.1cm}  \\
        \end{array}
        \right),
\end{equation}
where $\varPi_p p$ has the same dimension as $x$.

If a proper ordering technique is applied, the matrix $A$ from reservoir models
can be written as equation
\eqref{mat-ab},
\begin{equation}
\label{mat-ab}
{A} = \left(
        \begin{array}{lll}
        A_{pp} \hspace{0.1cm}   & A_{ps} \hspace{0.1cm} & A_{pw}   \\
        A_{sp} \hspace{0.1cm}   & A_{ss} \hspace{0.1cm} & A_{sw}   \\
        A_{wp} \hspace{0.1cm}   & A_{ws} \hspace{0.1cm} & A_{ww}   \\
        \end{array}
        \right),
\end{equation}
where the sub-matrix $A_{pp}$ is the matrix corresponding to the pressure unknowns, the
sub-matrix $A_{ss}$ is
the matrix corresponding to the
other unknowns, the sub-matrix $A_{ww}$ is the matrix corresponding to the well bottom hole pressure unknowns,
and other matrices are coupled items.

Let us introduce some notations for the preconditioning system $A x = f$.
If $A$ is a positive definite matrix, then we define the notation
$\mathscr{M}_g(A)^{-1} b$ to represent the solution $x$ from AMG methods.
If it is solved by the RAS method, then we use the notation $\mathscr{R}(A)^{-1}b$ to represent solution $x$.
The CPR-like preconditioners we develop are shown by
Algorithm \ref{pc-pcpr} to Algorithm \ref{pc-cpr-ffpf}, 
which are noted as CPR-FP, CPR-PF, CPR-FPF and CPR-FFPF methods \cite{cpr-chen}, respectively.

\begin{algorithm}[!htb]
\caption{The CPR-FP preconditioner}
\label{pc-pcpr}
\begin{algorithmic}[1]
\State $x = \mathscr{R}(A)^{-1} f $.
\State $r = f - Ax$
\State $x = x + \varPi_p (\mathscr{M}_g(A_{pp})^{-1} (\varPi_r  r))$.
\end{algorithmic}
\end{algorithm}

\begin{algorithm}[!htb]
    \caption{The CPR-PF preconditioner}
    \label{pc-pcpr-pf}
    \begin{algorithmic}[1]
        \State $x = \varPi_p (\mathscr{M}_g(A_{pp})^{-1} (\varPi_r  f))$
        \State $x = \mathscr{R}(A)^{-1} f $.
        \State $r = f - Ax$
        \State $x = x + \mathscr{R}(A)^{-1} f$.
    \end{algorithmic}
\end{algorithm}

\begin{algorithm}[!htb]
\caption{The CPR-FPF preconditioner}
\label{pc-cpr}
\begin{algorithmic}[1]
\State $x = \mathscr{R}(A)^{-1} f $.
\State $r = f - Ax$
\State $x = x + \varPi_p (\mathscr{M}_g(A_{pp})^{-1} (\varPi_r  r))$.
\State $r = f - Ax$
\State $x = x + \mathscr{R}(A)^{-1} r $.
\end{algorithmic}
\end{algorithm}

\begin{algorithm}[!htb]
    \caption{The CPR-FFPF preconditioner}
    \label{pc-cpr-ffpf}
    \begin{algorithmic}[1]
        \State $x = \mathscr{R}(A)^{-1} f $.
        \State $r = f - Ax$
        \State $x = x + \mathscr{R}(A)^{-1} r$.
        \State $r = f - Ax$
        \State $x = x + \varPi_p (\mathscr{M}_g(A_{pp})^{-1} (\varPi_r  r))$.
        \State $r = f - Ax$
        \State $x = x + \mathscr{R}(A)^{-1} r $.
    \end{algorithmic}
\end{algorithm}

\begin{figure}[!htb]
\centering
\begin{evb}
typedef struct CPR_PARS_
{
    RAS_PARS    ras;
    BMAMG_PARS  amg;
    INT         pres_which;
    INT         pres_loc;

    INT         itr_ras_pre;
    INT         itr_ras_post;

} CPR_PARS;

typedef struct CPR_DATA_
{
    RAS_DATA        ras;
    BMAMG_DATA      amg;

    INT             *pro_pres;    /* prolongation from pressure */
    INT             num_pro_pres;
    VEC             *varbuf;      /* vector r (A), buffer */
    VEC             *vpbbuf;      /* buffer for AMG */
    VEC             *vpxbuf;      /* buffer for AMG */

    CPR_PARS        pars;

} CPR_DATA;
\end{evb}
\caption{Data structure of CPR preconditioners}
\label{fig-ds-cpr}
\end{figure}

The data structure of the CPR preconditioners is shown in Figure \ref{fig-ds-cpr}. It has a RAS solver (\verb|ras|)
an AMG solver (\verb|amg|), prolongation information (\verb|pro_pres|) and \verb|num_pro_pres|), buffers and settings
for RAS solver and AMG solver.
The term \verb|CPR_PARS| stores parameters of the CPR methods.

\subsection{Data Structure for Preconditioners}

\begin{figure}[!htb]
\centering
\begin{evb}
/* pc type */
typedef enum PC_TYPE_
{
    PC_RAS,           /* Restricted Additive Schwarz */
    PC_AMG,           /* BommerAMG */
    PC_CPR_FP,        /* cpr */
    PC_CPR_PF,        /* cpr */
    PC_CPR_FPF,       /* cpr */
    PC_CPR_FFPF,      /* cpr */
    PC_USER,          /* user define */
    PC_NON,           /* no preconditioner */

} PC_TYPE;

/* preconditioner interface */
typedef void (*PC_USER_ASSEMBLE)(struct SOLVER_PC_ *pc, MAT *mat);
typedef void (*PC_SOLVE)(struct SOLVER_PC_ *pc, VEC *x, VEC *b);
typedef void (*PC_DESTROY)(struct SOLVER_PC_ *pc);

/* SOLVER_PC struct */
typedef struct SOLVER_PC_
{
    struct SOLVER_       *solver;     /* pointer to solver */

    void                 *data;
    PC_USER_ASSEMBLE     user_assemble;

    /* MPI */
    int                  rank;
    int                  nprocs;
    MPI_Comm             comm;

    PC_SOLVE             solve;
    PC_DESTROY           destroy;
    BOOLEAN              assembled;

} SOLVER_PC;
\end{evb}
\caption{Data structure of preconditioners}
\label{fig-ds-pc}
\end{figure}

The data structure for preconditioners is defined by Figure \ref{fig-ds-pc}. 
It provides three function pointers
that can complete assembling (\verb|PC_USER_ASSEMBLE|), solving (\verb|PC_SOLVE|)
and destroying (\verb|PC_DESTRORY|) a preconditioning system. With these function pointers,
this data structure is general purpose, and if a new set of implementations are provided, 
a new preconditioner can be
constructed. From the data structure, we can see built-in preconditioners, 
including RAS method, AMG methods and CPR methods,
are provided, and users can implement their own preconditioners by providing proper assembling, solving and destroying functions.
The data structure also contains a pointer to solver, and MPI info.

\section{Numerical Experiments}
\label{sec-exp}

A Blue Gene/Q from IBM that located in the IBM Thomas J. Watson Research Center
is employed. The system uses PowerPC A2 processor. Each processor has 18 cores and 16 cores are used
for computation.  Performance of each core is really low compared with processors from Intel.
However, it has a strong network relative to compute performance and the system is scalable.
Since the platform is designed for parallel applications, scalability is the most important objective.
In the following section, we will focus on scalability.

\subsection{Grid Partitioning}
\begin{figure}[!htb]
\begin{center}
\includegraphics[width=0.9\hsize, viewport=41 235 597 315, clip]{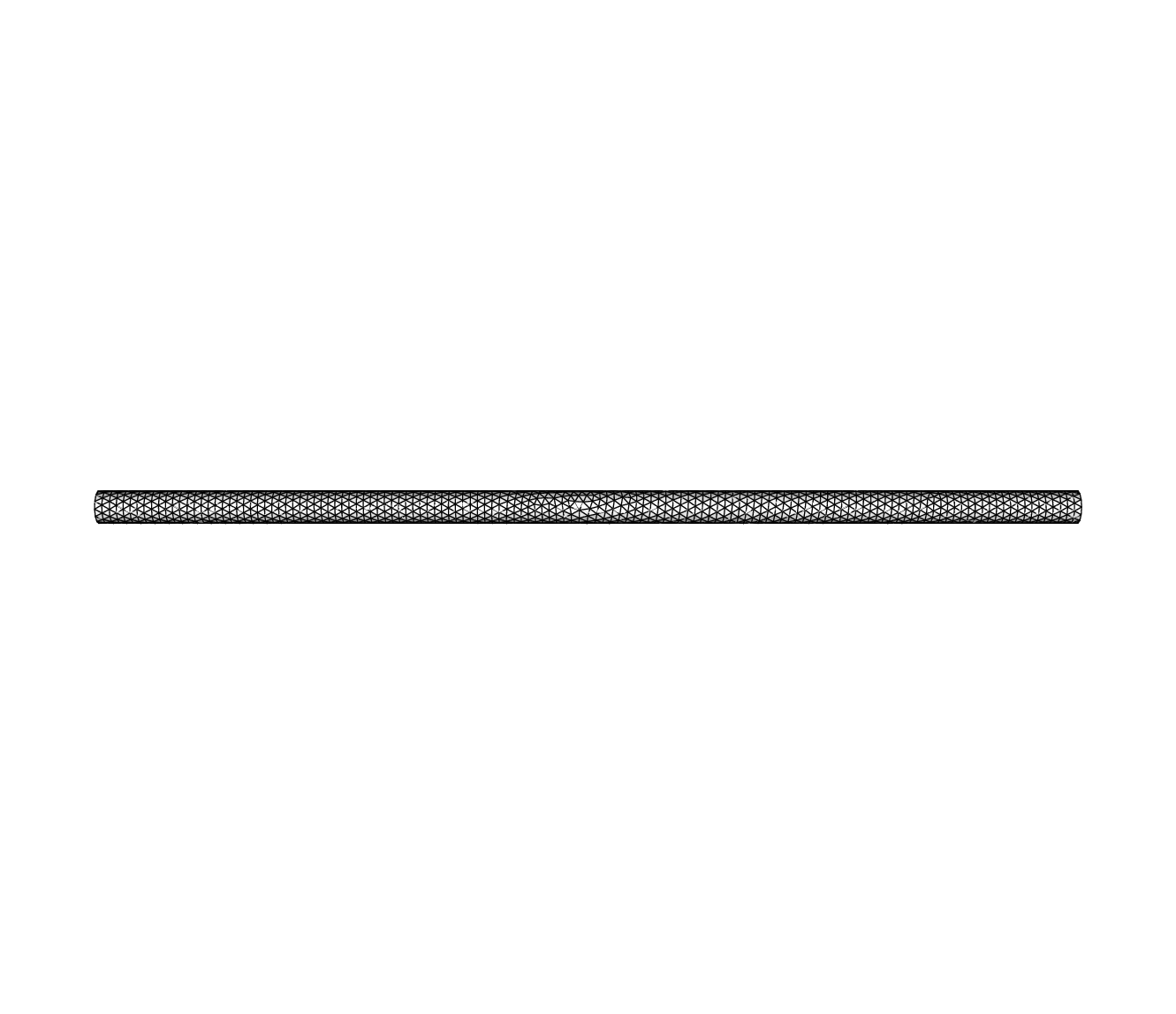}
\end{center}
\caption{Grid $\Omega_1$}
\label{fig-dlb-mesh1}
\end{figure}

\begin{figure}[!htb]
\centering
\includegraphics[width=0.6\hsize, viewport=118 75 520 476, clip]{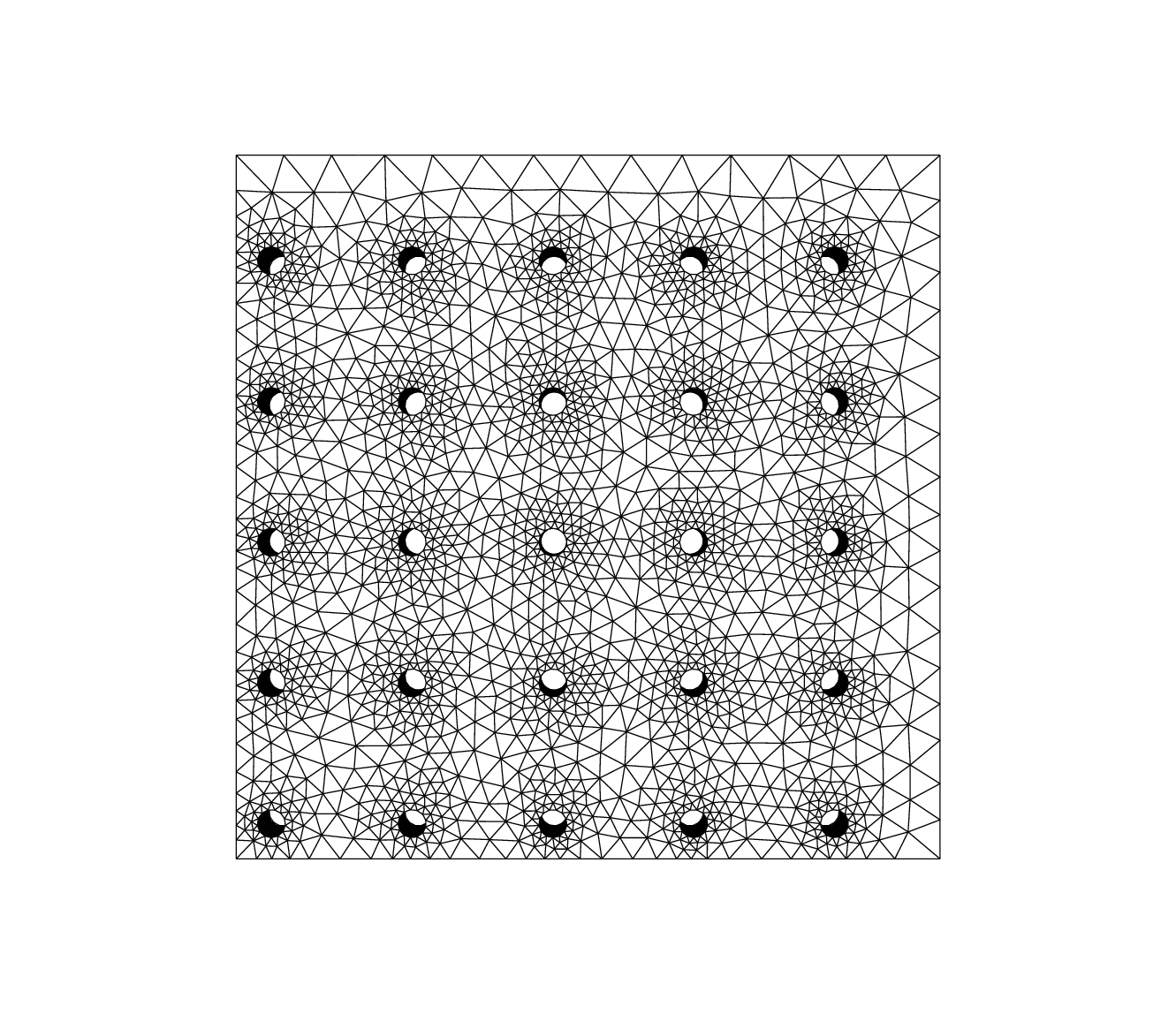}
\caption{Grid $\Omega_2$}
    \label{fig-dlb-mesh2}
\end{figure}

%%%%%%%%%%%%%%%%%%%%%%%%%%%%%%%%%%%%%%%%%%%%%%%%%%%%%%%%%%%%%%%%%%%%%%%%%%%%%%%%%%%%%%%%%%%%%%%
\begin{example}
    Two grids are applied to test the partitioning quality of the Hilbert space-filling
    curve method (HSFC), compared with other methods, such as the HSFC method from Zoltan,
    RCB (recursive coordinate bisection) from Zoltan, the
    Morton space-filling curve method, and ParMETIS. The first grid is
    $\Omega_1$, which is a long cylinder and contains 2,522,624 cells. The second one
    is a thin plate with many holes, which has 3,713,792 cells. They are shown
    in Figures \ref{fig-dlb-mesh1} and \ref{fig-dlb-mesh2}. Numerical results
    are presented in Tables \ref{tab-dlb-ic} and \ref{tab-dlb-si} \cite{lhths}.
    These grids are used for the finite element method and numerical experiments
    are from PHG (Parallel Hierarchical Grid) \cite{phg-quad}.
\end{example}

\begin{table}[!htb]
\centering \caption{Maximal inter-processor connectivities: $\Omega_1$ and $\Omega_2$}
    \label{tab-dlb-ic}
\begin{tabular}{|l|c|c|c|c|c|c|} \hline
\multicolumn{7}{|c|}{$\Omega_1$} \\ \hline
\# submeshes & 16 & 32 & 64 & 128&160 & 192   \\ \hline
ParMETIS     & 2  & 2  & 5  & 8  & 8  & 9 \\
RCB   & 2 & 2 & 5 & 11 & 12 & 13 \\
MSFC  & 4 & 5 & 10 & 16 & 19 & 21 \\
HSFC     & 3  & 6  & 13 & 23 & 24 & 23 \\
Zoltan/HSFC  & 12 & 18 & 21 & 23 & 24 & 24 \\
\hline
\multicolumn{7}{|c|}{$\Omega_2$} \\ \hline
\# submeshes &16  & 32 & 64 & 128& 160& 192   \\ \hline
ParMETIS     & 7  & 7  & 7  & 10 & 11 & 12 \\
RCB  & 7 & 7 & 8 & 11 &12 & 12 \\
MSFC &   9 & 13 & 18 & 21 &22 & 25 \\
HSFC     & 8  & 10 & 13 & 18 & 20 & 21 \\
Zoltan/HSFC  & 12 & 19 & 23 & 25 & 27 & 31 \\
\hline
\end{tabular}
\end{table}

\begin{table}[!htb]
\centering \tabcolsep=5pt \caption{Surface indices for $\Omega_1$ }
\label{tab-dlb-si}
\begin{tabular}{|l|c|c|c|c|c|c|} \hline
\multicolumn{7}{|c|}{maximum surface index (\%, $\Omega_1$)} \\ \hline
\# submeshes &16    & 32   & 64   & 128  & 160  & 192   \\ \hline
ParMETIS     & 0.74 & 1.48 & 2.88 & 5.00 & 5.78 & 5.71 \\
RCB & 1.0 & 2.18 & 4.07 & 6.06 & 7.06 & 7.60 \\
MSFC & 4.2 & 6.51 & 10.6 & 16.2 & 18.9 & 20.7 \\
HSFC     & 3.80 & 8.31 & 15.5 & 19.5 & 20.4 & 21.5 \\
Zoltan/HSFC  & 13.9 & 19.1 & 25.4 & 28.1 & 30.9 & 35.8 \\

\hline
\multicolumn{7}{|c|}{average surface index (\%, $\Omega_1$)} \\ \hline
\# submeshes &16    & 32   & 64   & 128  & 160  & 192   \\ \hline
ParMETIS     & 0.64 & 1.28 & 2.39 & 3.62 & 4.05 & 4.42 \\
RCB & 0.85 & 1.68 & 3.43 & 5.17 & 6.03 & 6.58 \\
MSFC & 2.93 & 5.14 & 7.44 & 9.8 & 10.7 & 11.5 \\
HSFC     & 2.78 & 5.03 & 7.18 & 9.34 & 10.2 & 10.8 \\
Zoltan/HSFC  & 8.72 & 11.8 & 16.2 & 20.6 & 22.0 & 23.6 \\

\hline
\end{tabular}
\end{table}

\begin{table}[!htb]
\centering \tabcolsep=5pt \caption{Surface indices for $\Omega_2$}
\label{tab-dlb-si2}
\begin{tabular}{|l|c|c|c|c|c|c|} \hline
\multicolumn{7}{|c|}{maximum surface index (\%, $\Omega_2$)} \\ \hline
\# submeshes &16    & 32   & 64   & 128  & 160  & 192   \\ \hline
ParMETIS     & 2.45 & 2.56 & 3.30 & 5.32 & 6.13 & 6.10 \\
RCB & 2.57 & 4.74 & 6.29 & 9.16 & 9.66& 11.6 \\
MSFC & 3.0 & 5.11 & 7.06 & 9.74 & 11.1 & 11.8 \\
HSFC     & 2.86 & 4.74 & 7.03 & 9.78 & 10.6 & 11.3 \\
Zoltan/HSFC  & 8.89 & 12.3 & 16.5 & 19.5 & 21.2 & 22.8 \\

\hline
\multicolumn{7}{|c|}{average surface index (\%, $\Omega_2$)} \\ \hline
\# submeshes &16    & 32   & 64   & 128  & 160  & 192   \\ \hline
ParMETIS     & 1.12 & 1.45 & 2.18 & 3.43 & 3.87 & 4.21 \\
RCB & 1.89 & 3.04 & 3.93 & 5.90 & 6.58 & 7.2 \\
MSFC & 2.24 & 3.61 & 5.21 & 7.34 & 8.22 & 8.92 \\
HSFC     & 2.19 & 3.29 & 4.88 & 6.92 & 7.81 & 8.27 \\
Zoltan/HSFC  & 5.10 & 6.86 & 9.47 & 12.6 & 14.1 & 15.3 \\

\hline
\end{tabular}
\end{table}

ParMETIS is a package for a graph method and other methods are geometric methods. In Table
\ref{tab-dlb-ic}, the inter-processor connectivity models the number of startups in group
communication, which measures latency. A smaller index means less latency, which is
good for parallel computing. We can see that ParMETIS is the best and the RCB method has
similar quality as ParMETIS. Other methods are worse than these two methods.

Tables \ref{tab-dlb-si} and \ref{tab-dlb-si2} present surface indices, which model a
communication volume. A smaller value means less communication. These two tables demonstrate again
that the graph methods can minimize communication. The results show that the RCB method has better
partitioning quality than other space-filling curve methods. However, our HSFC method
has better quality than the HSFC method from the Zoltan package.

%%%%%%%%%%%%%%%%%%%%%%%%%%%%%%%%%%%%%%%%%%%%%%%%%%%%%%%%%%%%%%%%%%%%%%%%%%%%%%%%%%%%%%%%%%%%%%%
\begin{example}
    \label{dlb-ex2}
    The computational domain is $1200 ft \times 2200 ft \times 170 ft$ and the grid
    size is $180 \times 660 \times 255$, which has 30,294,000 cells. The HSFC method
    and ParMETIS are applied to test partitioning quality. Numerical results are shown
    in Tables \ref{tab-dlb2-ic} and \ref{tab-dlb2-si}.
\end{example}

\begin{table}[!htb]
    \centering \caption{Maximal inter-processor connectivities, Example \ref{dlb-ex2}}
    \label{tab-dlb2-ic}
\begin{tabular}{|l|c|c|c|c|} \hline
    \multicolumn{5}{|c|}{Example \ref{dlb-ex2}} \\ \hline
\# submeshes &256    & 512   & 1024   & 2048  \\ \hline
ParMETIS     & 19  & 20  & 21 & 26  \\
HSFC         & 16  & 15  & 15 & 16  \\
\hline
\end{tabular}
\end{table}

\begin{table}[!htb]
    \centering \tabcolsep=5pt \caption{Surface indices for Eaxmple \ref{dlb-ex2}}
\label{tab-dlb2-si}
\begin{tabular}{|l|c|c|c|c|} \hline
\multicolumn{5}{|c|}{maximum surface index (\%)} \\ \hline
\# submeshes &256    & 512   & 1024   & 2048  \\ \hline
ParMETIS     & 6.86 & 9.23 & 11.0 & 13.6 \\
HSFC         & 8.34 & 10.7 & 13.5 & 18.2  \\

\hline
\multicolumn{5}{|c|}{average surface index (\%)} \\ \hline
\# submeshes &256    & 512   & 1024   & 2048  \\ \hline
ParMETIS     & 5.15 & 6.55 & 8.37 & 10.5  \\
HSFC         & 6.39 & 8.16 & 10.4 & 13.0  \\

\hline
\end{tabular}
\end{table}

Table \ref{tab-dlb2-ic} shows inter-processor connectivity. We can see that in this case
the HSFC method has better inter-processor connectivity, and the resulting partitioning has
less latency. Table \ref{tab-dlb2-si} shows that partitioning from the HSFC method has
more communication than ParMETIS.

%%%%%%%%%%%%%%%%%%%%%%%%%%%%%%%%%%%%%%%%%%%%%%%%%%%%%%%%%%%%%%%%%%%%%%%%%%%%%%%%%%%%%%%%%%%%%%%
\begin{example}
    \label{dlb-ex1}
    This example solves the \texttt{Helmholtz} problem using an adaptive finite element method,
    which has the \texttt{Dirichlet} boundary:
\[
\left \{
\begin{aligned}
-\Delta u + u  = &  ~f,  & (x, y, z) \in \Omega, \\
      u(x,y,z)   = &  ~g  & \text{ on } \partial\Omega.
\end{aligned}
\right.
\]
The computation domain is $\Omega_1$, and the equation has the analytical solution:
\[
u = \cos(2 \pi x)\cos(2 \pi y)\cos(2 \pi z).
\]
The initial grid has 4,927 cells (elements). The error indicator is defined by
\[ \eta_K^2 = h_K^2\|f_h + \Delta
u_h + u_h\|_{L^2(K)}^2 + \sum_{f \subset \partial
K\cap\Omega}{\frac{h_f}{2} \|[\frac{\partial u_h}{\partial
n_f}]\|^2_{L^2(f)}}.
\]
    Several partitioning methods are compared, including the HSFC (PHG/HSFC) method,
    the HSFC method from Zoltan (Zoltan/HSFC), RCB, ParMETIS, and the
    refinement-tree method (RTK).
\end{example}

\begin{figure}[!htb]
  \centering
  \includegraphics[width=0.85 \linewidth]{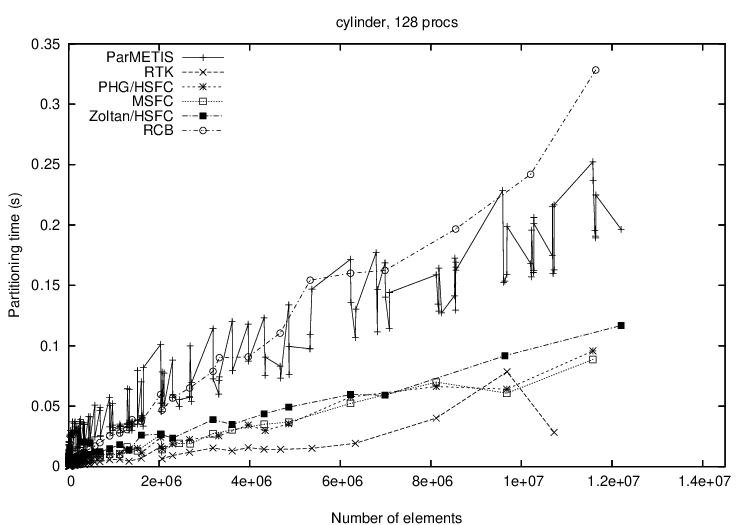}
  \caption{Partitioning time, Example \ref{dlb-ex1}}
  \label{cyl-rpt128}
\end{figure}

\begin{figure}[!htb]
  \centering
  \includegraphics[width=0.85 \linewidth]{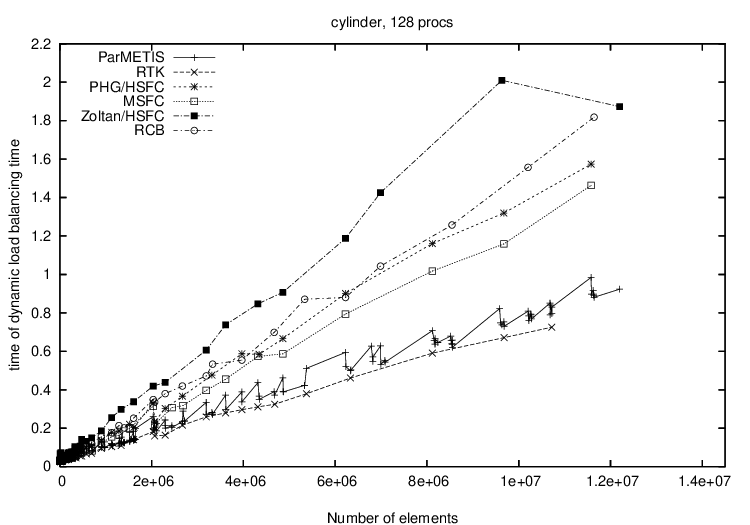}
  \caption{Dynamic load balancing time, Example \ref{dlb-ex1}}
  \label{cyl-dlb128}
\end{figure}

This example studies partitioning time and dynamic load balancing time. The dynamic load balancing
includes several stages, such as grid partitioning, data (DOF, degrees of freedom) exchange,
and sub-grid assembly. An adaptive finite element method is applied and the final grid size
is around 12 million cells (elements). Figure \ref{cyl-rpt128} shows that the refinement tree
method is the fastest method. The space-filling curve methods have a similar speed. The ParMETIS is
slower than these space-filling curve methods, but its speed is similar to the RCB method.
Figure \ref{cyl-dlb128} shows that the RTK method uses the least dynamic load balancing time and
is the fastest method, followed by ParMETIS. The Morton space-filling curve method has a similar speed
to our HSFC method. The RCB method is slightly slower but it is faster than the HSFC method from the Zoltan
package.

%%%%%%%%%%%%%%%%%%%%%%%%%%%%%%%%%%%%%%%%%%%%%%%%%%%%%%%%%%%%%%%%%%%%%%%%%%%%%%%%%%%%%%%%%%%%%%%
\begin{example}
    \label{dlb-ow}
    This case tests a two-phase oil-water model with a refined SPE10 project,
    where each original cell is refined into 27 smaller cells. The model
    has around 30 millions of cells. The nonlinear method is an inexact Newton method.
    The solver is GMRES(50) and the preconditioner is the CPR-PF method \cite{cpr-chen}.
    Only five time steps are applied. The case is run on GPC.
    Numerical summaries are shown in Table \ref{tab-ow-num}.
\end{example}

\begin{table}[!htb]
    \centering \tabcolsep=5pt
    \caption{Numerical summaries for Eaxmple \ref{dlb-ow}}
\label{tab-ow-num}
\begin{tabular}{|l|c|c|c|c|} \hline
\multicolumn{5}{|c|}{\# Newton iterations} \\ \hline
\# MPIs &256    & 512   & 1024   & 2048  \\ \hline
ParMETIS     & 106 & 105 & 105 & 79 \\
HSFC         & 101 & 80 & 105 & 101  \\

\hline
\multicolumn{5}{|c|}{\# Linear iterations} \\ \hline
\# MPIs &256    & 512   & 1024   & 2048  \\ \hline
ParMETIS     & 364 & 231 & 332 & 375  \\
HSFC         & 251 & 289 & 379 & 296  \\

\hline

\multicolumn{5}{|c|}{Overall running time (s)} \\ \hline
\# MPIs &256    & 512   & 1024   & 2048  \\ \hline
ParMETIS     & 668.5 & 346.6 & 236.1 & 116.4  \\
HSFC         & 621.0 & 288.4 & 295.2 & 187.2  \\

\hline
\multicolumn{5}{|c|}{Average time for each Newton iteration (s)} \\ \hline
\# MPIs &256    & 512   & 1024   & 2048  \\ \hline
ParMETIS   & 6.30 & 3.30 & 2.25 & 1.47 \\
HSFC       & 6.15 & 3.61 & 2.81 & 1.85  \\

\hline
\end{tabular}
\end{table}

The CPR-PF method is a combination of the algebraic multi-grid method (AMG) and the
restricted additive Schwarz method (RAS) \cite{RAS}. The grid partitioning
affects the distribution of matrices and these two numerical methods.
Table \ref{tab-ow-num} shows that the number of Newton iterations varies form 79 to 106
under different settings. The linear iterations are from 231 to 379. The
overall running time shows that when using 256 and 512 MPIs, the simulations with the HSFC method
are faster, and when using 1024 and 2048 MPIs, the simulations with ParMETIS
are faster. Considering the average running time for each Newton iteration, the simulations
with ParMETIS use less time than simulations with the HSFC method.

\begin{example}
    \label{dlb-ogw}
    This case tests a three-phase problem with a refined SPE10 geological model,
    where each cell is refined into 27 smaller cells. The model
    has around 30 millions of cells. The nonlinear method is an inexact Newton method.
    The solver is GMRES(30) and the preconditioner is the CPR-PF method \cite{cpr-chen}.
    Only five time steps are applied. The case is run on GPC again.
    Numerical summaries are shown in Table \ref{tab-ogw-num}.
\end{example}

\begin{table}[!htb]
    \centering \tabcolsep=5pt \caption{Numerical summaries for Eaxmple \ref{dlb-ogw}}
\label{tab-ogw-num}
\begin{tabular}{|l|c|c|c|c|} \hline
\multicolumn{5}{|c|}{\# Newton iterations} \\ \hline
\# MPIs &256    & 512   & 1024   & 2048  \\ \hline
ParMETIS     & 22 & 23 & 19 & 23 \\
HSFC         & 22 & 22 & 22 & 22  \\

\hline
\multicolumn{5}{|c|}{\# Linear iterations} \\ \hline
\# MPIs &256    & 512   & 1024   & 2048  \\ \hline
ParMETIS     & 141 & 145 & 110 & 149  \\
HSFC         & 146 & 132 & 142 & 142  \\

\hline

\multicolumn{5}{|c|}{Overall running time (s)} \\ \hline
\# MPIs &256    & 512   & 1024   & 2048  \\ \hline
ParMETIS     & 456.8 & 259.2 & 138.3 & 249.6  \\
HSFC         & 441.4 & 234.3 & 143.4 & 118.8  \\

\hline
\multicolumn{5}{|c|}{Average time for each Newton iteration (s)} \\ \hline
\# MPIs &256    & 512   & 1024   & 2048  \\ \hline
ParMETIS   & 20.7 & 11.3 & 7.28 & 10.85 \\
HSFC       & 20.1 & 10.7 & 6.52 & 5.4  \\

\hline
\end{tabular}
\end{table}

This example shows different behavior from Example \ref{dlb-ow}. Table \ref{tab-ogw-num}
shows that the simulations with the HSFC partitioning method are faster than the simulations with ParMETIS.

\subsection{SpMV}

\begin{figure}[!htb]
    \centering
    \includegraphics[width=0.5\linewidth, angle=270]{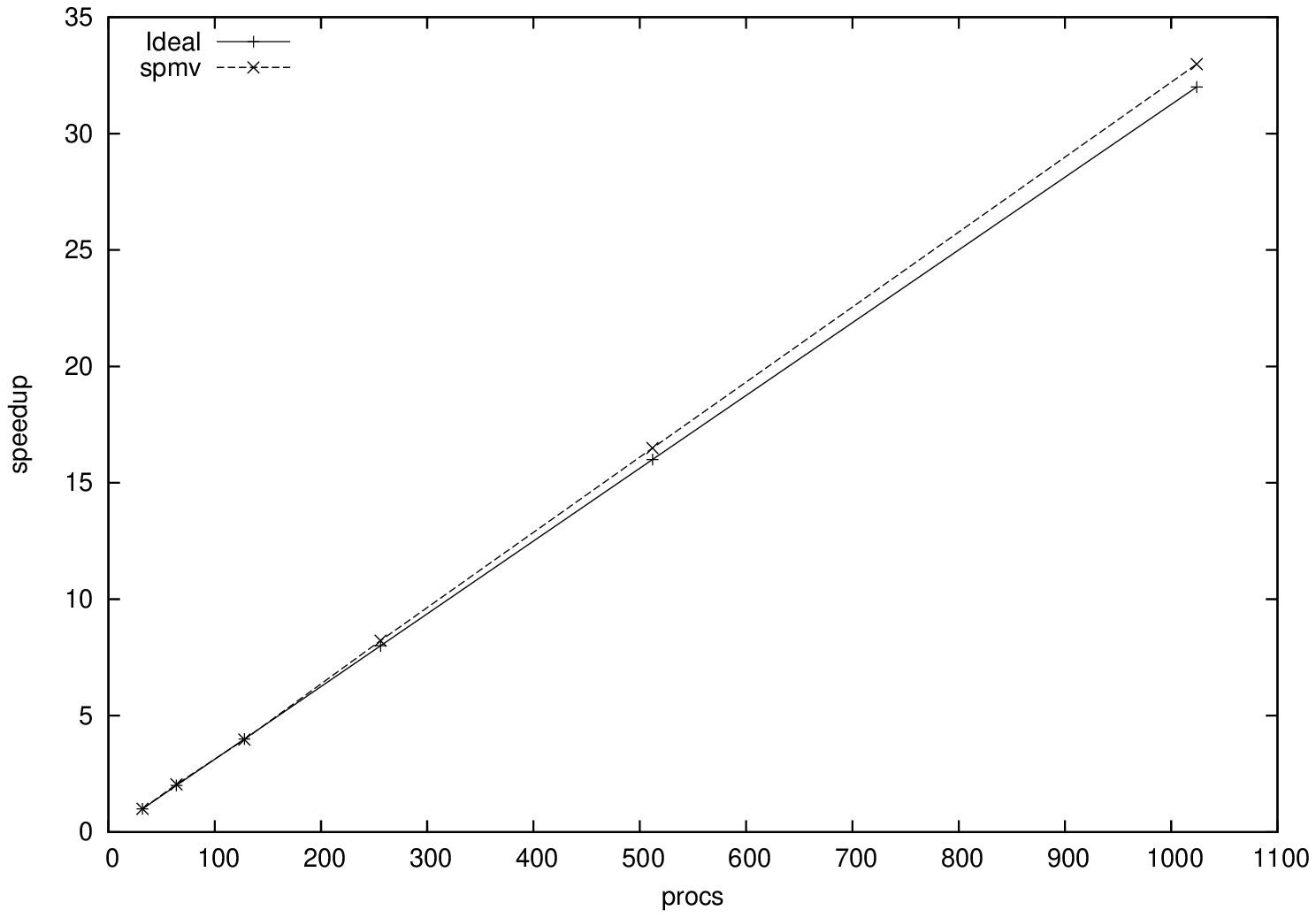}
    \caption{Scalability of SpMV, Example \ref{ex-spmv}}
    \label{fig-spmv}
\end{figure}

\begin{example}
\label{ex-spmv}
This example tests the performance of sparse matrix-vector multiplication (SpMV) on IBM Blue Gene/Q.
The matrix is a square matrix from a Poisson equation and it has 200 millions rows.
Its performance is shown in Table \ref{tab-spmv} and its scalability is presented in
Figure \ref{fig-spmv}.
\end{example}

\begin{table}[!htb]
  \centering
  \begin{tabular}{|c|c|c|c|c|c|c|} \hline
  \# procs & 32    & 64    & 128   & 256    & 512     & 1024      \\ \hline
  Time (s) & 2.211 & 1.078 & 0.556 & 0.269  & 0.134   & 0.067     \\ \hline
  \end{tabular}
  \caption{Performance of SpMV, Example \ref{ex-spmv}}
  \label{tab-spmv}
\end{table}

This example uses up to 128 compute cards, when more than 128 MPI tasks are used,
multiple MPI tasks run on one card. From Table \ref{tab-spmv}, we can see that when
MPI tasks are doubled, the running time of SpMV is reduced by half. This example
show our SpMV kernel has excellent scalability.
Speedup is compared with ideal condition in Figure \ref{fig-spmv}, which shows that our SpMV kernel
has good scalability.

\subsection{Poisson Equation}

\begin{figure}[!htb]
    \centering
    \includegraphics[width=0.5\linewidth, angle=270]{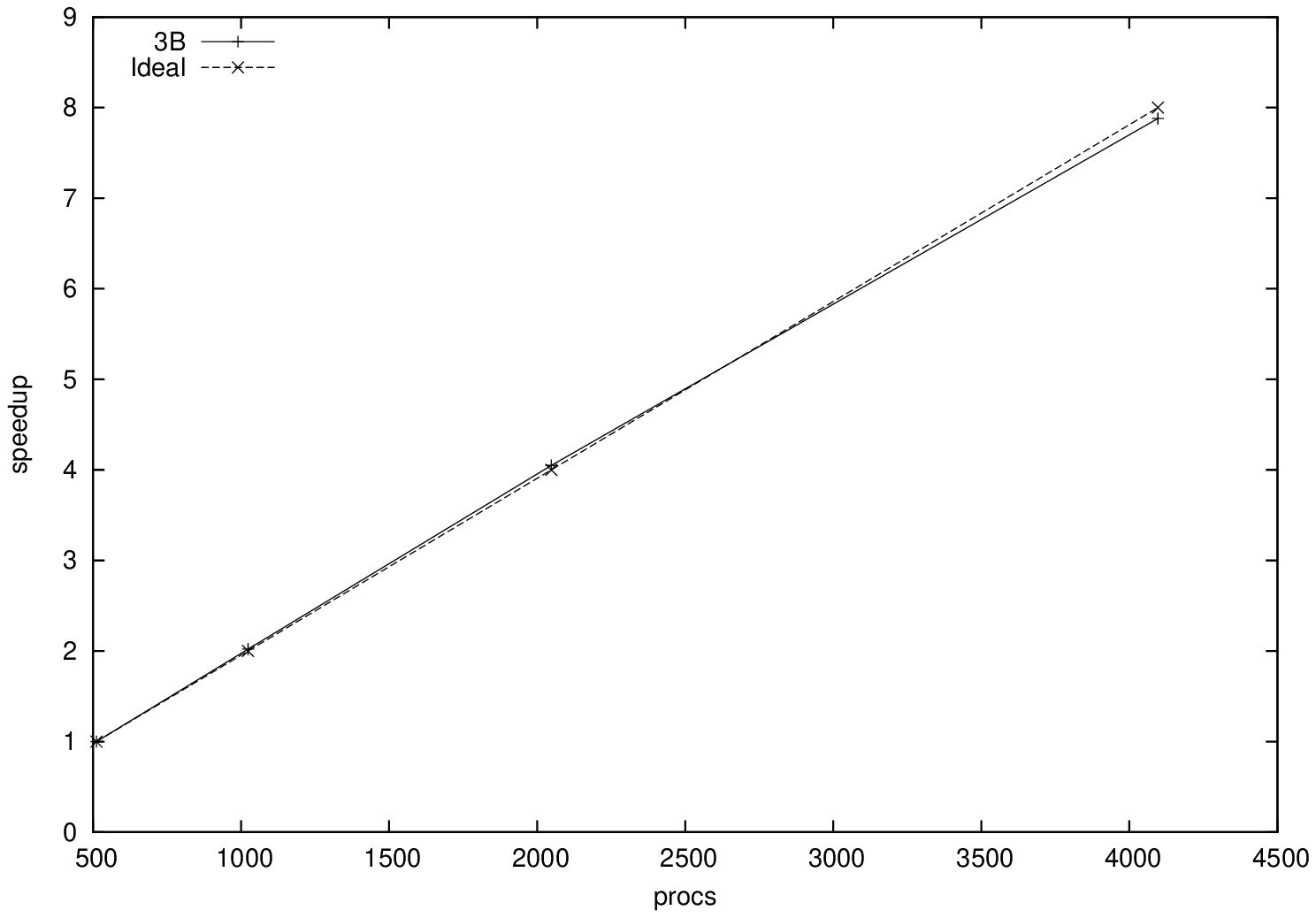}
    \caption{Scalability of Example \ref{ex-216}}
    \label{fig-ex216}
\end{figure}

\begin{example}
\label{ex-216}
This example tests a Poisson equation with 3 billions of grid cells. The example uses
up to 4,096 CPU cores (MPIs). The linear solver is GMRES(30) method with RAS preconditioner.
The solver runs 90 iterations. The overlap of RAS method is one, and sub-domain problem
on each core is solved by ILU(0).
The numerical summaries are reported in Table \ref{tab-ex216} and scalability results
are shown in Figure \ref{fig-ex216}.
\end{example}

\begin{table}[!htb]
  \centering
  \begin{tabular}{cccccc} \hline
  \# procs & Gridding & Building & Assemble & Overall (s)  & Speedup  \\ \hline
  512   & 217.0   & 29.16   & 66.42 & 918.91 & 1.0      \\
  1024  & 98.83   & 14.79   & 33.71 & 454.04 & 2.02     \\
  2048  & 47.53   & 7.49    & 17.47 & 227.05 & 4.05     \\
  4096  & 23.31   & 3.86    & 9.17  & 116.64 & 7.88     \\ \hline
  \end{tabular}
  \caption{Numerical summaries of Example \ref{ex-216}}
  \label{tab-ex216}
\end{table}

Table \ref{tab-ex216} shows numerical summaries of Poisson equation. This example tests strong scalability
of our platform. Here the gridding time includes grid generation, grid partitioning and grid redistribution. From
Table \ref{tab-ex216}, we can see that the gridding has excellent scalability and when MPIs are doubled, running
time for gridding reduces by half. Building time is the time spent on generation of a linear system $Ax = b$. Since
there is no communication involved, the scalability is ideal. Assemble time includes time for generating linear solver,
and time for generating preconditioner (RAS method). The overall time includes gridding time, building time,
assemble time and solution time. Again, from Table \ref{tab-ex216} and Figure \ref{fig-ex216}, we can see
our platform has excellent scalability.

\begin{figure}[!htb]
    \centering
    \includegraphics[width=0.5\linewidth, angle=270]{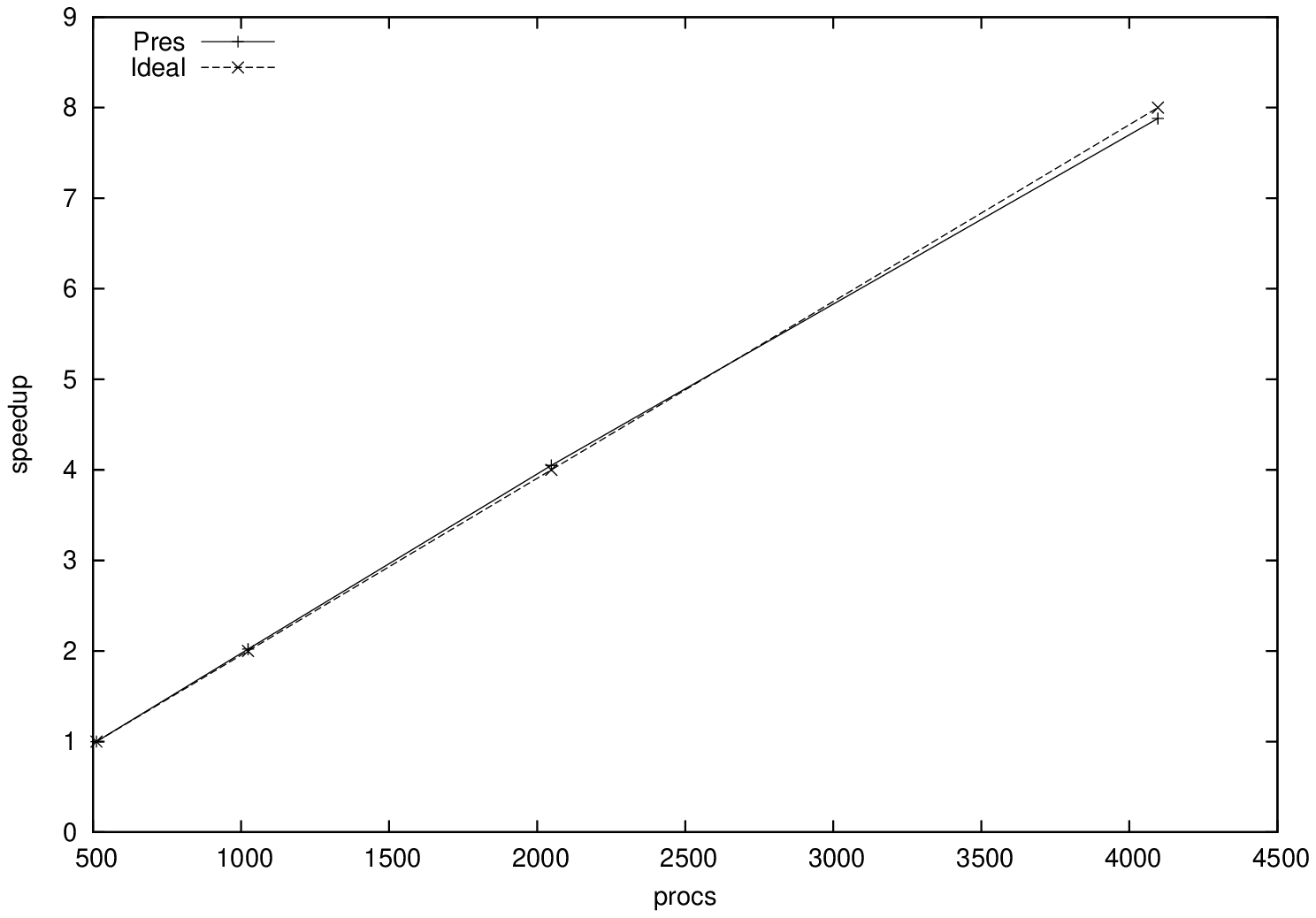}
    \caption{Scalability of the large-scale model}
    \label{fig-po-bgq}
\end{figure}

\begin{example}
\label{ex-6b}
The case goes with the size of the matrix at 6 billion variables and is constructed from 
    pressure equations. GMRES(30) was applied to solve the system with RAS 
    (Restricted Additive Schwarz) as the preconditioner and fixed iterations at 90. 
    IBM Blue Gene/Q was used to carry the simulation. 
    Numerical summaries are shown in Table \ref{tab-po-bgq}
    and scalability is presented in Fig \ref{fig-po-bgq}.
\end{example}

\begin{table}[!htb]
\centering
  \caption{Numerical summaries of the large-scale model}
\begin{tabular}{ccccc} \\ \hline
  \# procs  & Gridding (s) & Build (s) & Assemble (s)  & Overall (s) \\ \hline
  512 & 429.78 & 57.83 & 130.54 & 1829.18 \\
  1024 & 200.24 & 29.28 & 66.12 & 906.73   \\
  2048&  106.83 & 14.83 & 34.03&  463.59\\
  4096 & 47.82 & 7.59 & 17.7  & 232.77  \\
 \hline
\end{tabular}
  \label{tab-po-bgq}
\end{table}

This example tests the scalability of grid generation, building of linear system, and solution of
linear system (including solver, preconditioner and SpMV). Table \ref{tab-po-bgq}
shows that when MPI tasks are doubled, running time of grid generation, building of linear system,
and solution of linear system is cut by half, which means our platform and linear solvers have
excellent scalability. The scalability is demonstrated by
Figure \ref{fig-po-bgq} demonstrate. This example also show that the solver can calculate extremely
large-scale linear systems.

\subsection{Oil-water Model}

\begin{figure}[!htb]
    \centering
    \includegraphics[width=0.5\linewidth, angle=270]{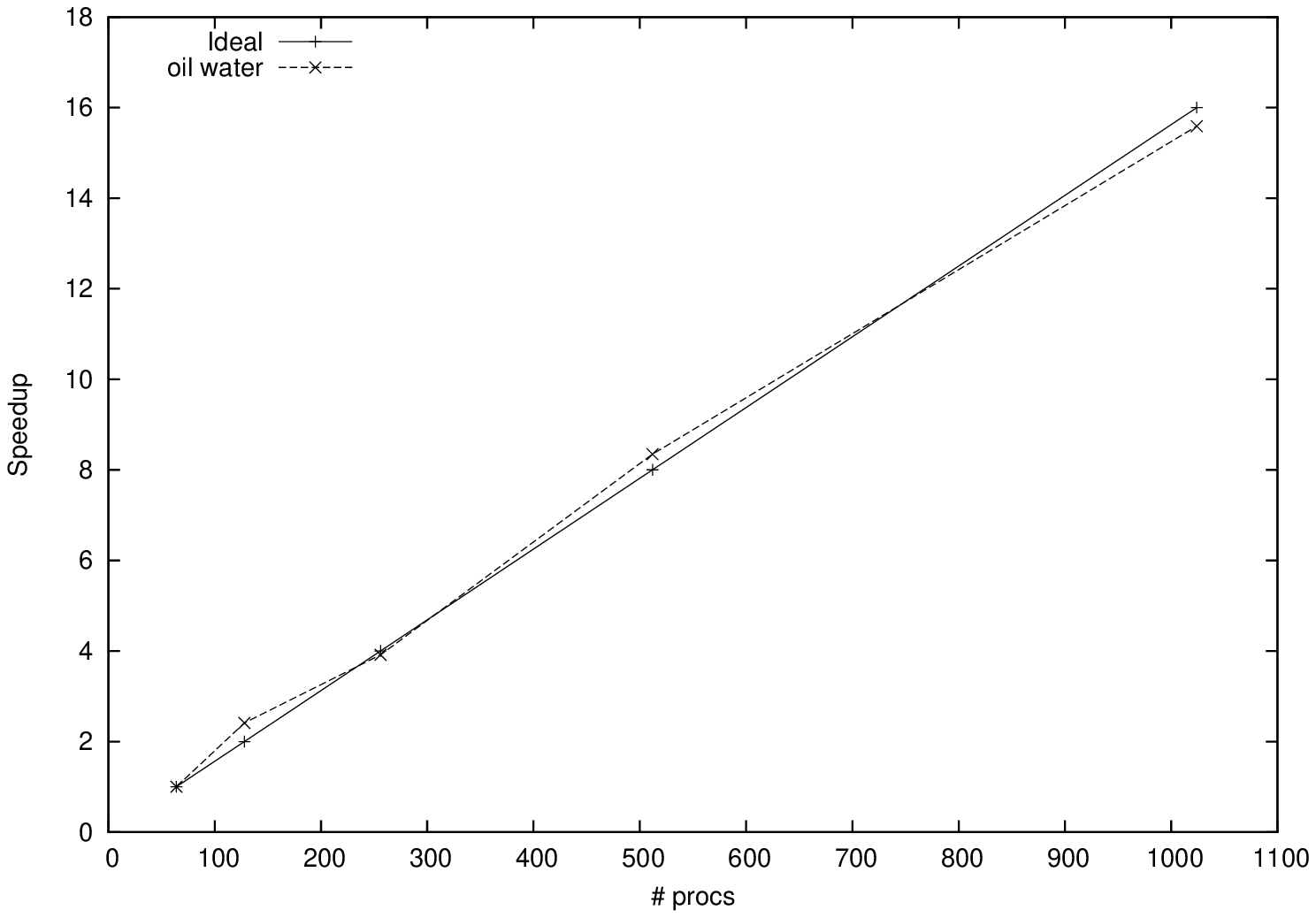}
    \caption{Scalability of Example \ref{ex-scal-ex4}}
    \label{fig-scal-e4}
\end{figure}

\begin{example}
\label{ex-scal-ex4}
This example tests a refined SPE10 case for the two-phase oil-water model, where each grid cell
is refined into 27 grid cells. This case has around 30 millions of grid cells and around 60 millions of unknowns.
The stopping criterion for the inexact Newton method is 1e-3 and the maximal Newton iterations are 20.
The BiCGSTAB solver is applied and its maximal iterations are 100. The preconditioner is the CPR-FPF preconditioner.
The potential reordering and the Quasi-IMPES decoupling strategy are applied.
The simulation period is 10 days. Up to 128 compute cards are used.
The numerical summaries are shown
in Table \ref{ex-scal-e4}, and the speedup (scalability) is shown in
Figure \ref{fig-scal-e4}.
\end{example}

\begin{table}[!htb]
\centering
\begin{tabular}{lccrcrc} \\ \hline
  \# procs   & \# Steps & \# Ntn & \# Slv & \# Avg-S  & Time (s) & Avg-T (s)\\ \hline
  64   & 50 & 315 & 3451 & 10.9 & 119167.4 & 378.3 \\
  128  & 48 & 286 & 3296 & 11.5 & 49488.7  & 173.0 \\
  256  & 54 & 323 & 4190 & 12.9 & 30423.2  & 94.1 \\
  512  & 52 & 329 & 3635 & 11.0 & 14276.5  & 43.3 \\
  1024 & 54 & 316 & 3969 & 12.5 & 7643.9   & 24.1 \\
\hline
\end{tabular}
  \caption{Numerical summaries of Example \ref{ex-scal-ex4}}
  \label{ex-scal-e4}
\end{table}

In this example, up to 1,024 MPI tasks are employed and the simulation with 64 MPI tasks
is used as the base case to calculate speedup and scalability. The numerical summaries in Table \ref{ex-scal-e4}
show the inexact Newton method is robust, where around 50 time steps and around 300 Newton iterations are
used for each simulation with different MPI tasks. The linear solver BiCGSTAB and the preconditioner CPR-FPF
show good convergence, where the average number of linear iterations for each nonlinear iteration is between 10 and 13.
The results mean our linear solver and preconditioner are effective and efficient.
The overall running time and average time for each Newton iteration show our simulator has excellent scalability
on IBM Blue Gene/Q, which is almost ideal for parallel computing. The scalability is also
demonstrated in Figure \ref{fig-scal-e4}. The running time and scalability curve also demonstrate
our linear solver and preconditioner are scalable for large-scale simulation.

\begin{figure}[!htb]
    \centering
    \includegraphics[width=0.5\linewidth, angle=270]{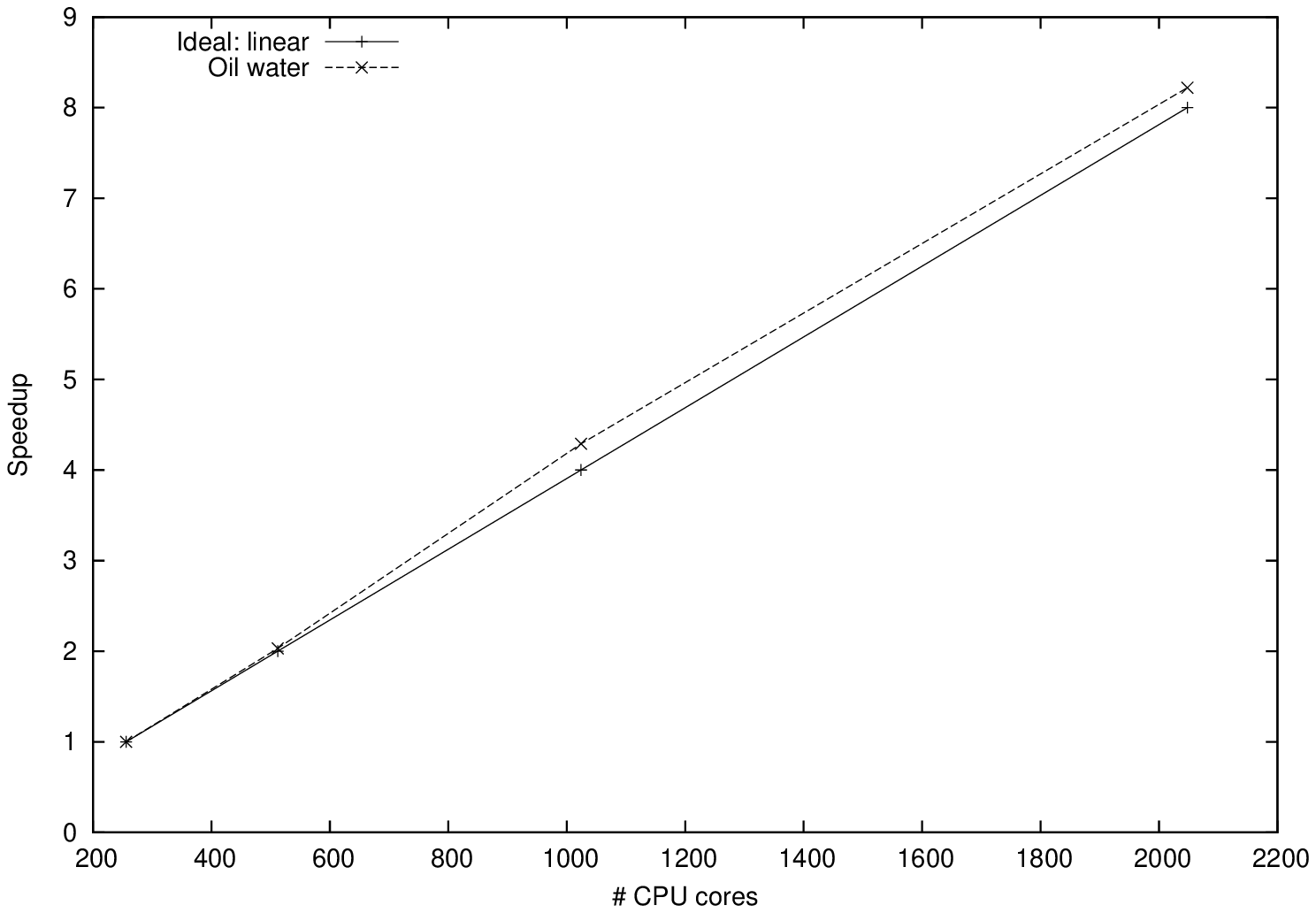}
    \caption{Scalability of Example \ref{ex-scal-ex42}}
    \label{fig-scal-e42}
\end{figure}

\begin{example}
\label{ex-scal-ex42}
This example tests a refined SPE10 case for the two-phase oil-water model, where each grid cell
is refined into 125 grid cells. This case has around 140 millions of grid cells and around 280 millions of unknowns.
The stopping criterion for the inexact Newton method is 1e-2 and the maximal Newton iterations are 20.
The GMRES(30) solver is applied and its maximal iterations are 100. The preconditioner is the CPR-FPF preconditioner.
The potential reordering and the Quasi-IMPES decoupling strategy are applied.
The simulation period is 2 days. Up to 128 compute cards are used.
The numerical summaries are shown in Table \ref{ex-scal-e42}, and the speedup (scalability) is shown in
Figure \ref{fig-scal-e42}.
\end{example}

\begin{table}[!htb]
\centering
\begin{tabular}{lccrcrc} \\ \hline
  \# procs   & \# Steps & \# Ntn & \# Slv & \# Avg-S  & Time (s) & Avg-T (s)\\ \hline
  256  & 36 & 225 & 5544 & 24.6 & 117288.1 & 521.2 \\
  512  & 36 & 226 & 5724 & 25.3 & 57643.1 & 255.0 \\
  1024 & 35 & 207 & 5446 & 26.3 & 27370.3 & 132.2 \\
  2048 & 36 & 209 & 5530 & 26.4 & 14274.9 & 68.3 \\
\hline
\end{tabular}
  \caption{Numerical summaries of Example \ref{ex-scal-ex42}}
  \label{ex-scal-e42}
\end{table}

In this example, up to 2048 MPI tasks are employed and the simulation with 256 MPI tasks
is used as the base case to calculate speedup and scalability. The numerical summaries in Table \ref{ex-scal-e42}
show the inexact Newton method is robust, where around 36 time steps and around 220 Newton iterations are
used for each simulation with different MPI tasks. The linear solver GMRES(30) and the preconditioner CPR-FPF
show good convergence, where the average number of linear iterations for each nonlinear iteration is around 26.
The overall running time and average time for each Newton iteration show our simulator has excellent scalability
on IBM Blue Gene/Q, which is almost ideal for parallel computing and shows slight super-linear scalability.
The scalability is also
demonstrated in Figure \ref{fig-scal-e42}. The results show
our linear solver and preconditioner are scalable for large-scale simulation.

\subsection{Dual Porosity Oil-water Model}

\begin{figure}[!htb]
    \centering
    \includegraphics[width=0.5\linewidth, angle=270]{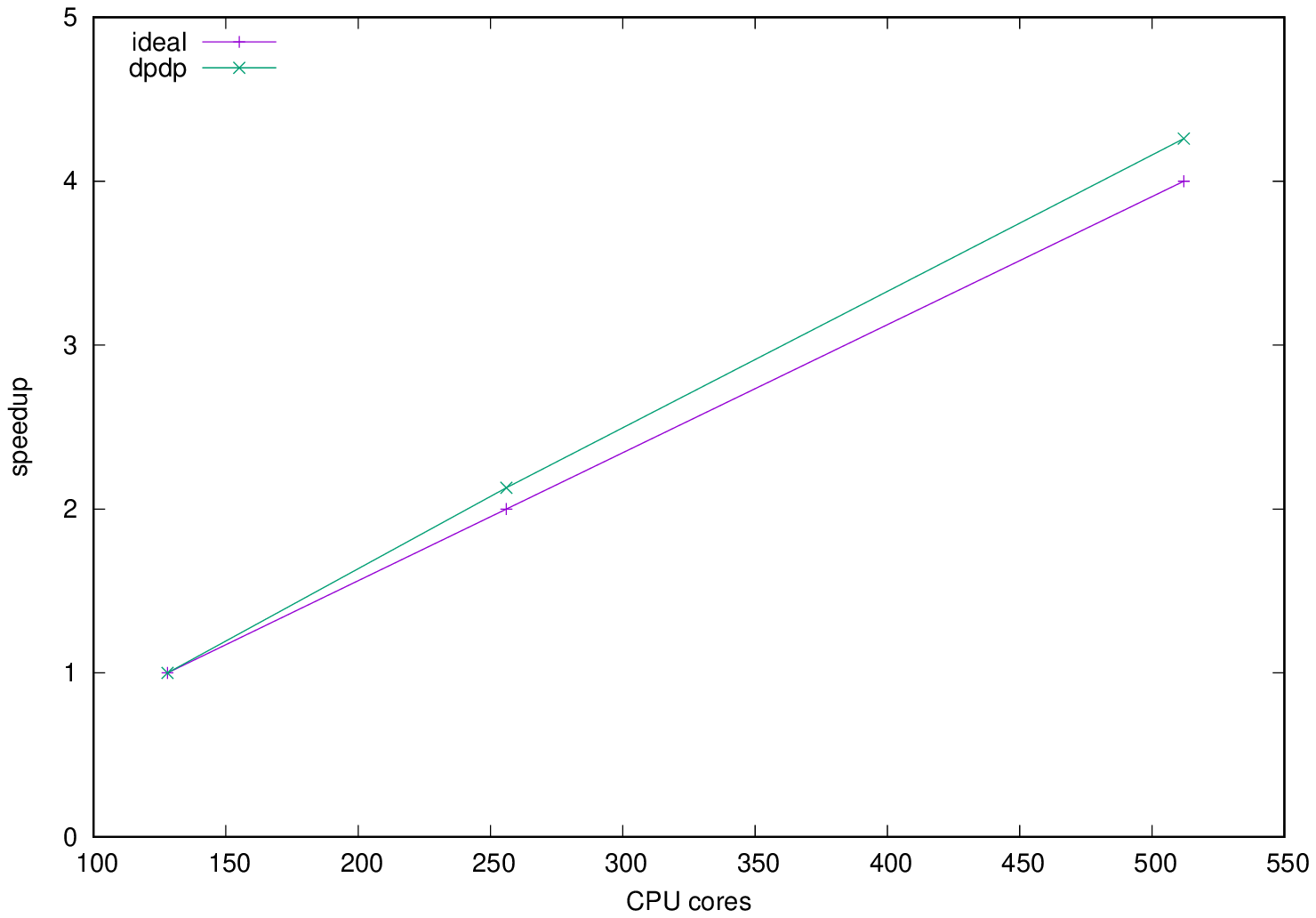}
    \caption{Scalability of Example \ref{dpdp-lg2}}
    \label{fig-dpdp-lg2}
\end{figure}

\begin{example}
    \label{dpdp-lg2}
    This example tests the scalability of a large model with 320 million grid cells, and 
    the dimension is $400\times 800\times 1000$. The reservoir has fractures, which is modelled by dual
    porosity method. The linear systems have around 1.3 billion unknowns.
    The simulation period is 30 days and its maximal time step is
    10 days. The non-linear system is solved by standard Newton method with a termination tolerance of 1e-4,
    The maximal Newton iterations are 20.
    And the linear systems are solved by BICGSTAB solver and CPR-FPF preconditioner. Its termination tolerance
    is 1e-3, and the maximal iterations are 100.
    The numerical summaries are listed in Table \ref{tab-dpdp-lg2} and scalability results
    are shown in Figure \ref{fig-dpdp-lg2}.
\end{example}

\begin{table}[!htb]
  \centering
  \begin{tabular}{cccccl} \hline
      \# procs   & Steps & \# Newton &  \# Solver & Time (s)   & Speedup  \\ \hline
      128        & 16       & 31     & 34         & 6471.55    & 1        \\
      256        & 16       & 31     & 35         & 3033.45    & 2.13     \\
      512        & 16       & 31     & 35         & 1517.12    & 4.26     \\
      \hline
  \end{tabular}
  \caption{Numerical summaries of Example \ref{dpdp-lg2}}
  \label{tab-dpdp-lg2}
\end{table}

Table \ref{tab-dpdp-lg2} presents numerical performance of the oil-water model with dual prosity method, which
models the natural fractured reservoirs. The table shows that the non-linear method, solver and the parallel
CPR-FPF preconditioner are effective and efficient. Excellent scalability is obtained, which is also confirmed
by Figure \ref{fig-dpdp-lg2}.

\subsection{Black Oil Model}
\begin{example}
\label{ex8}
The example tests the scalability of the black oil simulator using a refined SPE10 geological model,
where each grid cell is refined to 27 grid cells. The model has 30.3 millions
of grid cells. The inexact Newton method is applied and the termination tolerance is
$10^{-2}$. The linear solver is BiCGSTAB, whose maximal inner iterations are 100.
The preconditioner is the CPR-FPF method and the overlap for the RAS method is one.
The potential reordering and the ABF methods are enabled.
The simulation period is 10 days.
The maximal change allowed in one time step of pressure is 1,000 psi and
the maximal change of saturation is 0.2. Up to 128 compute cards are used.
Summaries of numerical results are shown in Table \ref{tab-ex8}.
\end{example}

\begin{table}[!htb]%[H]
  \centering
\begin{tabular}{cccrcr} \hline
\# procs  & \# Steps & \# Ntn & \# Slv & \# Avg-S  & Time (s)\\ \hline
64   &  33 & 292 & 1185 & 4.0 & 106265.9  \\
128  & 33 & 296 & 1150 & 3.8 & 50148.3  \\
256  & 33 & 299 & 1267 & 4.2 & 25395.8  \\
512  & 33 & 301 & 1149 & 3.8 & 12720.5  \\
1024 & 33 & 301 & 1145 & 3.8 & 6814.2 \\ \hline
\end{tabular}
  \caption{Numerical summaries of Example \ref{ex8}}
\label{tab-ex8}
\end{table}

\begin{figure}[!htb]
    \centering
    \includegraphics[width=0.5\linewidth, angle=270]{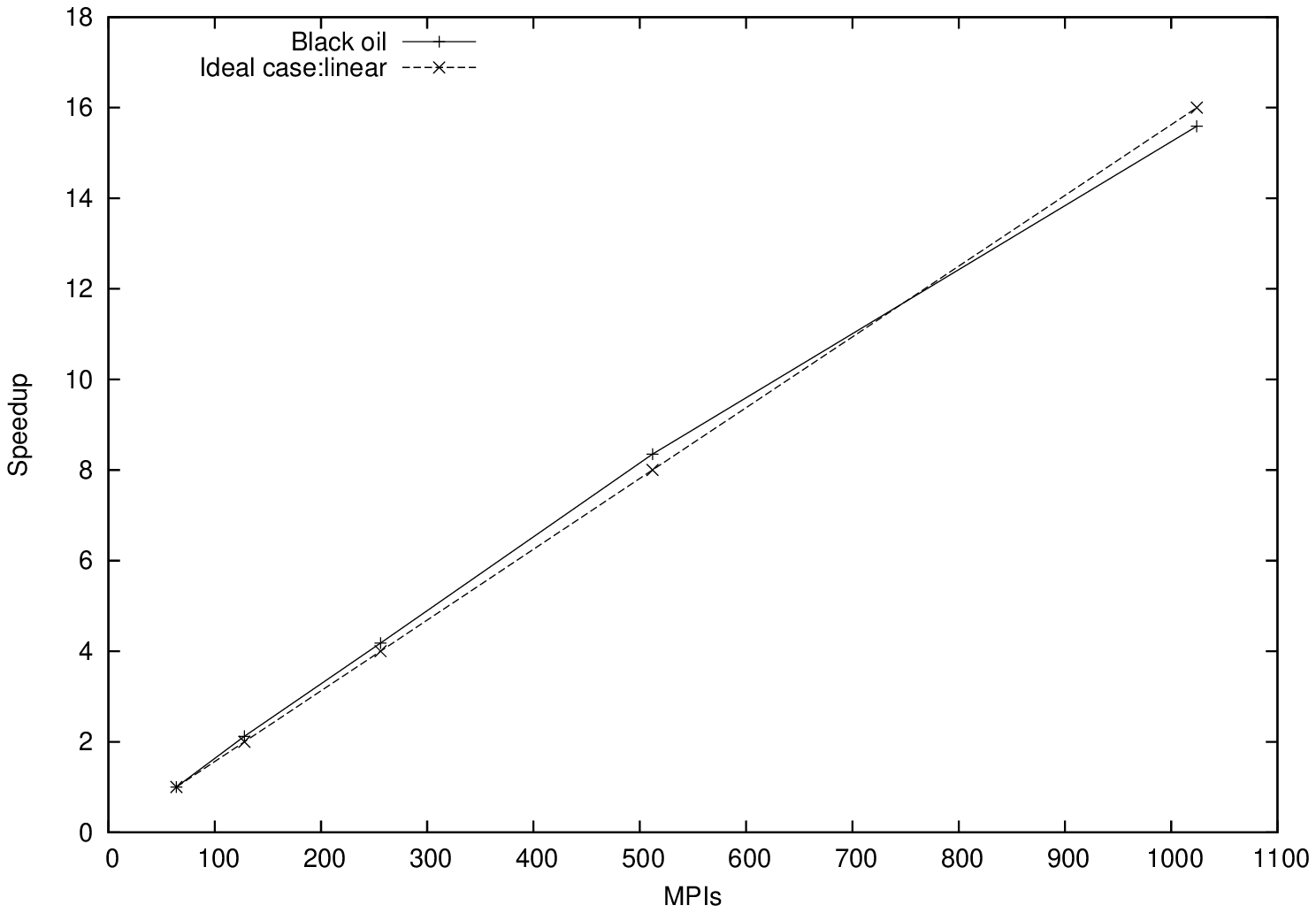}
    \caption{Scalability (speedup) of Example \ref{ex8}}
    \label{fig-ex8}
\end{figure}

Table \ref{tab-ex8} includes information for the nonlinear method, linear solver and running time. For all
simulations, 33 time steps are used and the total Newton iterations are around 300. The results show
the inexact Newton method is robust. For the linear solver and preconditioner, their convergence is good,
which terminate in around 4 iterations. The results mean the linear solver and preconditioner are
robust and effective for this highly heterogeneous model. The running time, average time per Newton
iteration and scalability curve in Figure \ref{fig-ex8} show the scalability of our simulator, linear solver
and preconditioner is good. When we use up to 1,024 MPI tasks and each compute card runs up to 8
MPI tasks, the scalability is excellent.

\begin{example}
\label{bos-1}
The case is a refined SPE1 project with 100 millions of grid cells. Linear solver is BiCGSTAB.
Potential reordering and ABF decoupling are applied.
Numerical summaries are in Table \ref{tab-bos-1}.
\end{example}

\begin{table}[!htb]
\centering
\begin{tabular}{lccrcr} \hline
  MPIs   & \# Steps & \# Newton & \# Solver & \# Avg. Itr  & Time (s)\\ \hline
    512  & 27 & 140 & 586 & 4.1 & 11827.9  \\
    1024 & 27 & 129 & 377 & 2.9 & 5328.4  \\
    2048 & 26 & 122 & 362 & 2.9 & 2708.5 \\
    4096 & 27 & 129 & 394 & 3.0 & 1474.2 \\
\hline
\end{tabular}
\caption{Numerical summaries, Example \ref{bos-1}}
\label{tab-bos-1}
\end{table}

\begin{figure}[!htb]
\begin{center}
    \includegraphics[width=0.5\linewidth, angle=270]{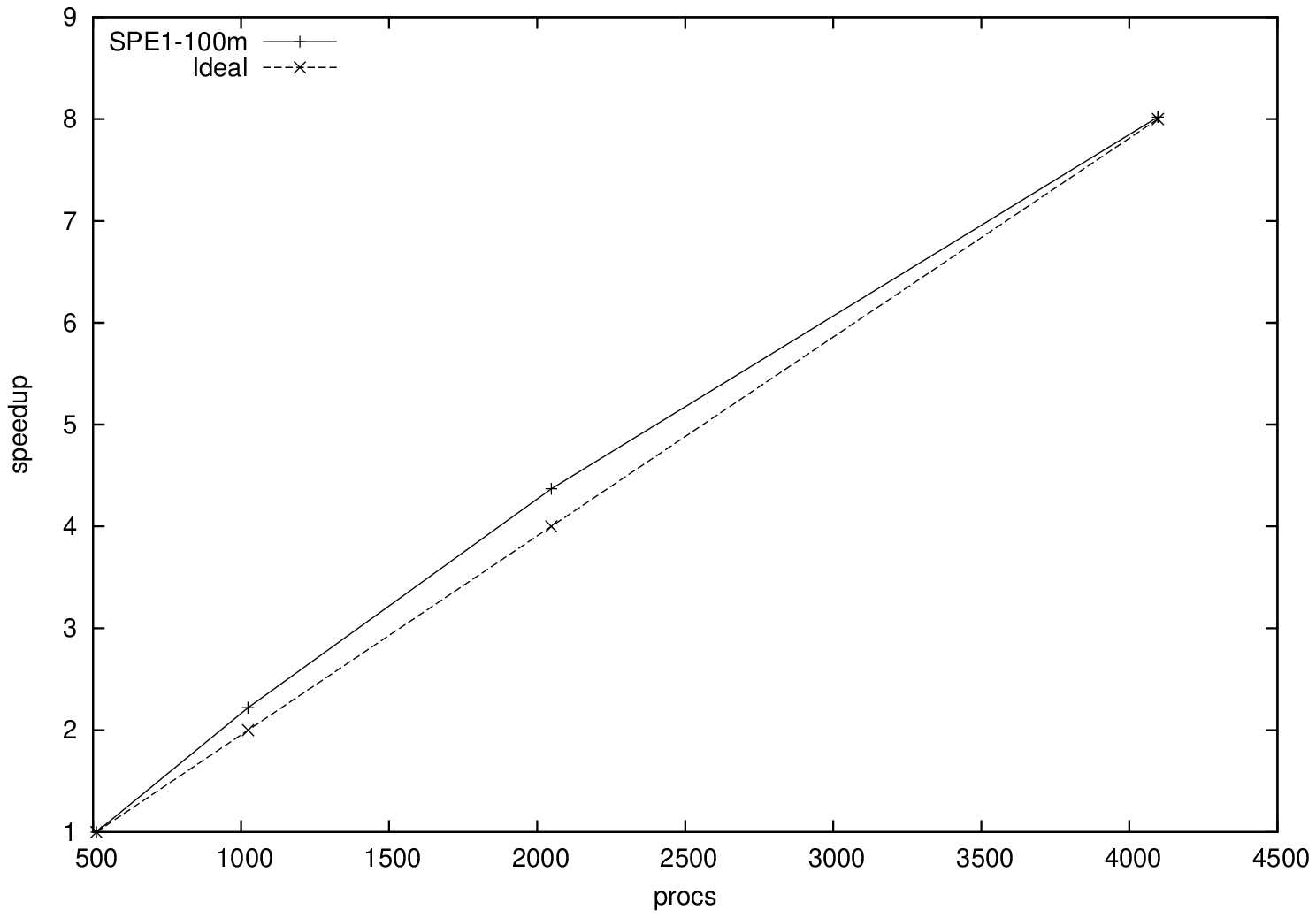}
\end{center}
\caption{Scalability of Example \ref{bos-1}}
\label{fig-bos1}
\end{figure}

\subsection{Thermal Model}

\begin{figure}[!htb]
    \centering
    \includegraphics[width=0.5\linewidth, angle=270]{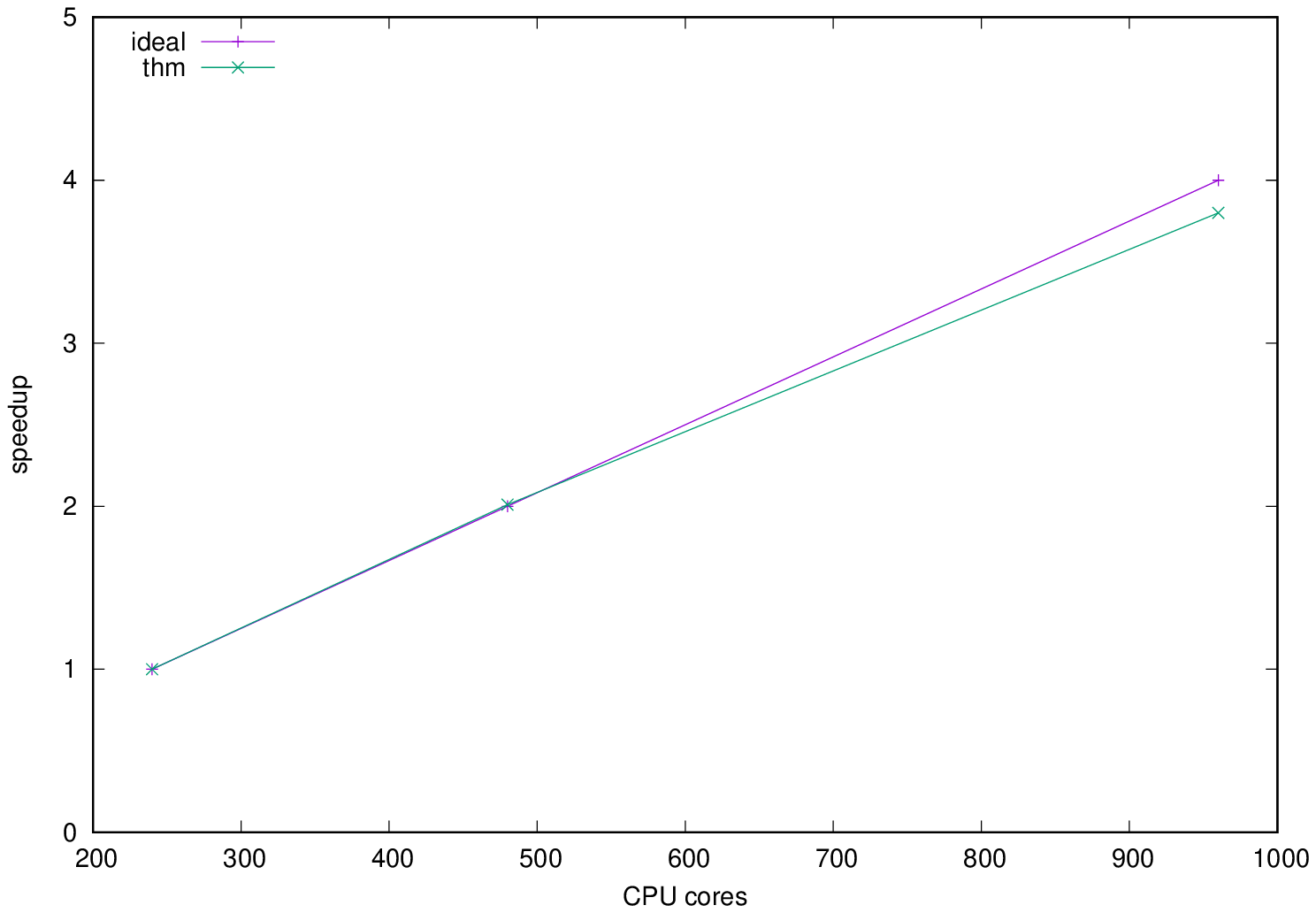}
    \caption{Scalability of Example \ref{thm-ex1}}
    \label{fig-thm-ex1}
\end{figure}

\begin{example}
    \label{thm-ex1}
    This example tests the scalability of a large model with 1.15 billion grid cells, and 
    the dimension is $360\times 2000\times 1600$. The linear systems have 4.6 billion unknowns.
    The simulation period is 5 time steps. The non-linear system is solved by standard Newton method with a
    termination tolerance of 1e-5,
    The maximal Newton iterations are 15.
    And the linear systems are solved by BICGSTAB solver and CPR-FPF preconditioner. Its termination tolerance
    is 1e-10, and the maximal iterations are 100.
    The numerical summaries are listed in Table \ref{tab-thm-ex1} and scalability results
    are shown in Figure \ref{fig-thm-ex1}.
\end{example}

\begin{table}[!htb]
  \centering
  \begin{tabular}{cccccl} \hline
      \# procs   & Steps & \# Newton &  \# Solver & Time (s)   & Speedup  \\ \hline
      240        & 5        & 5      & 5          & 1802.23    & 1        \\
      480        & 5        & 5      & 5          & 897.69     & 2.01     \\
      960        & 5        & 5      & 5          & 474.75     & 3.80     \\
      \hline
  \end{tabular}
  \caption{Numerical summaries of Example \ref{thm-ex1}}
  \label{tab-thm-ex1}
\end{table}

This model is a large-scale thermal model. Table \ref{tab-thm-ex1} shows the non-linear method is effective
and one Newton iteration is required for each time step. The linear solver and preconditioner is effective,
which could solve a linear system of order 4.6 billion in one iteration. The thermal simulator has excellent
scalability, which is demonstrated by the table and Figure \ref{fig-thm-ex1}.

\section{Conclusion}
Our work on developing an in-house parallel platform is presented in this paper, which
provides grids, data, linear solvers and preconditioners for reservoir simulators.
Various techniques and methods have been introduced, including the Hilbert space-filling curve method,
topological partitioning method, structured grid, distributed matrices and vectors, and
multi-state preconditioners for reservoir simulations. Examples,
including grid management, linear solvers, pressure equations, and 
simulators, are presented to benchmark our platform.
Numerical results show that our platform and simulators have excellent scalability
and applications based on the platform can be sped up thousands of times faster.
This paper also shows parallel computing is a powerful tool for large-scale
scientific computing.

\section*{Acknowledgement}
The support of Department of Chemical and Petroleum Engineering and Reservoir Simulation Group, University of
Calgary is gratefully acknowledged. The research is
partly supported by NSERC/AIEES/Foundation CMG and AITF Chairs.

\bibliographystyle{elsarticle-num}
\bibliography{<your-bib-database>}

\end{document}